# Rigid body motion in viscous flows using the Finite Element Method

M.I. Herreros[1,2*], S. Ligüérzana[3]

[1]Centro de Astrobiología (CSIC-INTA), Instituto Nacional de Técnica Aeroespacial, 28850 Torrejón de Ardoz, Madrid, Spain
[2]Departamento de Ingeniería Térmica y Fluidos, Universidad Carlos III de Madrid, 28911 Leganés, Madrid, Spain
[3]Agencia Estatal de Meteorología (AEMET), 28071 Madrid, Spain

*Corresponding author

## Abstract

A new model for the numerical simulation of a rigid body moving in a viscous fluid flow using FEM is presented. One of the most interesting features of this approach is the small computational effort required to solve the motion of the rigid body, comparable to a pure fluid solver. The model is based on the idea of extending the fluid velocity inside the rigid body and solving the flow equations with a penalty term to enforce rigid motion inside the solid. In order to get the velocity field in the fluid domain the Navier-Stokes equations for an incompressible viscous flow are solved using a fractional-step procedure combined with the two-step Taylor-Galerkin for the fractional linear momentum. Once the velocity field in the fluid domain is computed, calculation of the rigid motion is obtained by averaging translation and angular velocities over the solid. One of the main challenges when dealing with the fluid-solid interaction is the proper modelling of the interface which separates the solid moving mass from the viscous fluid. In this work the combination of the level set technique and the two-step Taylor-Galerkin algorithm for tracking the fluid-solid interface is proposed. The characteristics exhibited by the two-step Taylor-Galerkin, minimizing oscillations and numerical diffusion, make this method suitable to accurately advect the solid domain avoiding distortions at its boundaries, and thus preserving the initial size and shape of the rigid body. The proposed model has been validated against empirical solutions, experimental data and numerical simulations found in the literature. In all tested cases, the numerical results have shown to be accurate, proving the potential of the proposed model as a valuable tool for the numerical analysis of the fluid-solid interaction.

## 1. Introduction

Movement description of immersed bodies is present in several industrial processes (sedimentation procedures, metallurgical industry, etc.), but it is also of great interest in different scientific areas, such as geological sedimentation processes, asteroid impacts or fluvial transport and deposition, to name a few. In most of these cases, it becomes necessary the study of the interaction among a set of rigid bodies immersed in a fluid flow. To this end, and as a starting point, one single rigid body dynamics is usually analysed. Despite of the large number of works addressing this issue in the last years, development of new models capable to accurately deal with this problem is still a challenge nowadays.

Modelling the motion of an immersed body in a fluid flow requires the solution of the coupled fluid and solid dynamics. In addition to that, a procedure must be provided for the tracking of the solid position in time, preserving its volume and shape. Many efforts have been devoted in the last years to solve these issues.

In order to track the fluid-solid interface, some formulations have been proposed in the past within the frame of classical mesh-based methods. The deformable spatial domain [1] and the arbitrary Lagrangian-Eulerian formulations [2-4] intended to describe the body position by just moving the mesh. However, these methods presented some problems when a high relative motion among the bodies occurred. To circumvent this issue, an improved mesh updated method was introduced by Behr and Tezduyar [5]. Nevertheless, many alternative formulations were proposed in order to avoid the numerical inconveniences of remeshing. In this context the immersed boundary method [6-9] and the fictitious domain method [10-16] are especially remarkable. In the fictitious domain method, the fluid motion equations are extended inside the solid particles' domains. Lagrangian multipliers are defined in the rigid bodies domains to match over these regions the fluid flow and solid motion velocities. Some other formulations, however, combine techniques to describe the surface of the particles along with the Lattice Boltzmann method to solve the dynamics of the problem in a fixed mesh [17-20]. In addition, a wide variety of meshfree methods have been proposed over the last years: diffuse element method [21], element free Galerkin method [22-23], reproducing kernel particle method [24-25], h-p cloud method [26], partition of unity method [27], meshless local Petrov–Galerkin method [28], finite point method [29], radial basis function [30-31], Smoothed Particle Hydrodynamics (SPH) [32-38] or Taylor-SPH [39-43], among others. The main advantage of meshfree methods is that numerical solutions can be achieved without using a computational grid, thus dealing in a straightforward manner with particle tracking and therefore avoiding some of the difficulties encountered in classical mesh-based methods. However, most of these meshless methods present other numerical issues, such us smearing in their solutions, lack of consistency or problems when dealing with boundary conditions.

In this paper a numerical model for the direct computational simulation of freely moving rigid bodies in fluids using the Finite Element Method (FEM) is presented. The proposed computational algorithm is based on the idea of extending the fluid velocity inside the rigid body and solving the flow equations with a penalty term to enforce rigid motion inside the solid [15-16]. One of the most interesting features of this approach is that the low computational effort required to solve the motion of the rigid body is similar to that of a pure fluid solver [15]. In order to get the velocity field in the fluid domain, the Navier-Stokes equations for an incompressible viscous flow are solved using the fractional-step procedure proposed by Chorin [44] combined with the two-step Taylor-Galerkin [45-56] to solve the fractional linear momentum equation. Once the velocity field in the fluid domain is computed, calculation of the rigid motion is obtained by averaging translation and angular velocities over the solid. One of the main problems found in the fluid-solid interaction is the proper modelling of the interface which separates the solid moving mass from the viscous fluid. In this work the combination of the level set technique [16,45,57-61] and the two-step Taylor-Galerkin algorithm for tracking the fluid-solid interface is adopted [45-56]. The characteristics exhibited by the two-step Taylor-Galerkin [47-56], minimizing oscillations and numerical diffusion, make this method suitable to accurately advect the solid domain avoiding distortions at its boundaries, and thus preserving the initial size and shape of the rigid body.

The paper is structured as follows. First, the mathematical model and the proposed computational algorithm are presented in Section 2. Next, in Section 3, the Navier-Stokes equations are discretized, velocity in the solid domain is obtained and the rigid body properties are advected by means of the level set technique. To demonstrate the performance of the proposed method, some numerical examples are presented in Section 4, which is followed by the final conclusions in Section 5.

## 2. Mathematical model

Let us consider a rigid solid S($t$) submerged in an incompressible flow contained in a domain Ω. Thus F($t$) = Ω − S($t$) is the fluid domain. The fluid-solid interaction problem can be modelled by the Navier-Stokes equations in F($t$) and the rigid motion of the solid S($t$):

$$\rho\frac{\partial \mathbf{u}}{\partial t}+\rho\ div(\mathbf{u}\otimes\mathbf{u}) = div\ \boldsymbol{\sigma}+\rho\mathbf{g} \qquad \text{for } \mathbf{x}\in \mathrm{F}(t) \text{ and } t>0 \tag{1}$$

$$\bar{\mathbf{u}} = \mathbf{V}+\mathbf{W}\times(\mathbf{x}-\mathbf{x}_G) \qquad \text{for } \mathbf{x}\in \mathrm{S}(t) \text{ and } t>0 \tag{2}$$

$$M\dot{\mathbf{V}} = -\int_\Sigma \boldsymbol{\sigma}(\mathbf{u},p)\ \mathbf{n}\ d\Omega + M\mathbf{g} \tag{3}$$

$$\mathbf{I}\dot{\mathbf{W}} = -\int_\Sigma \boldsymbol{\sigma}(\mathbf{u},p)\ \mathbf{n}\times(\mathbf{x}-\mathbf{x}_G)\ d\Omega \tag{4}$$

where Σ($t$) is the fluid-solid interface, $\mathbf{n}$ is the normal pointing towards the solid, $\mathbf{g}$ is gravity; $\rho(\mathbf{x},t)$ and $p(\mathbf{x},t)$ are the density and pressure fields; $\mathbf{u}(\mathbf{x},t)$ and $\bar{\mathbf{u}}(\mathbf{x},t)$ are the velocities in the fluid and solid domains; and $M$, $\mathbf{x}_G$ and $\mathbf{I}$ are the mass, center of gravity and inertia matrix of the solid, respectively.

Equations (3) and (4) translate the solid acceleration as a result of gravity and fluid forces at the interface, where the linear and angular velocities are noted as $\mathbf{V}$ and $\mathbf{W}$. The stress tensor $\boldsymbol{\sigma}$ is defined as

$$\sigma_{ij}(\mathbf{u},p) = \frac{\mu}{\rho}\left(\frac{\partial u_i}{\partial x_j}+\frac{\partial u_j}{\partial x_i}\right)-p\delta_{ij} \tag{5}$$

being $\mu$ the dynamic viscosity of the fluid and $\delta_{ij}$ the Kronecker delta.

The above system has to be complemented by the initial and boundary conditions in Ω and the Dirichlet boundary condition on the fluid-solid interface Σ($t$).

## 2.2. Computational algorithm

The proposed computational algorithm is based on Patankar´s projection method [15-16]. The idea consists of extending the fluid velocity inside the solid body and solving the flow equations with a penalty term to enforce rigid motion inside the solid. Considering a penalty parameter $\lambda \gg 1$, the momentum equation applicable in the entire domain Ω can be written as:

$$\rho(\mathbf{x},t)\frac{\partial \mathbf{u}}{\partial t}+\rho(\mathbf{x},t)div(\mathbf{u}\otimes\mathbf{u}) = div\boldsymbol{\sigma}+\rho(\mathbf{x},t)\mathbf{g}+\lambda\mathcal{H}(\mathbf{x},t)\rho(\mathbf{x},t)(\bar{\mathbf{u}}-\mathbf{u}) \tag{6}$$

coupled with the incompressibility condition:

$$div\ \mathbf{u} = 0 \tag{7}$$

being:

$$\mathcal{H}(\mathbf{x},t) = \begin{cases} 1 & if\quad \mathbf{x}\in \mathrm{S}(t) \\ 0 & if\quad \mathbf{x}\in \mathrm{F}(t) \end{cases}$$

The penalty method described in (6)-(7) is proved to converge to the solution of the original problem (1)-(4) in the limit $\lambda\to\infty$ [62-63]. Therefore, Dirichlet boundary conditions at the fluid-solid interface, Σ($t$), are no longer required since all the geometric information is now included in the $\mathcal{H}$ function. The penalization approach proposed herein is physically motivated by the idea that a solid boundary can be modelled as a permeable solid with a very small permeability, $1/\lambda \rightarrow 0$ [64].

In the proposed model, the density field, $\rho(\boldsymbol{x},t)$, can be expressed in terms of an indicator function, $\phi(\boldsymbol{x},t)$, which identifies the portion of the domain occupied by either the solid or the fluid (Figure 1).

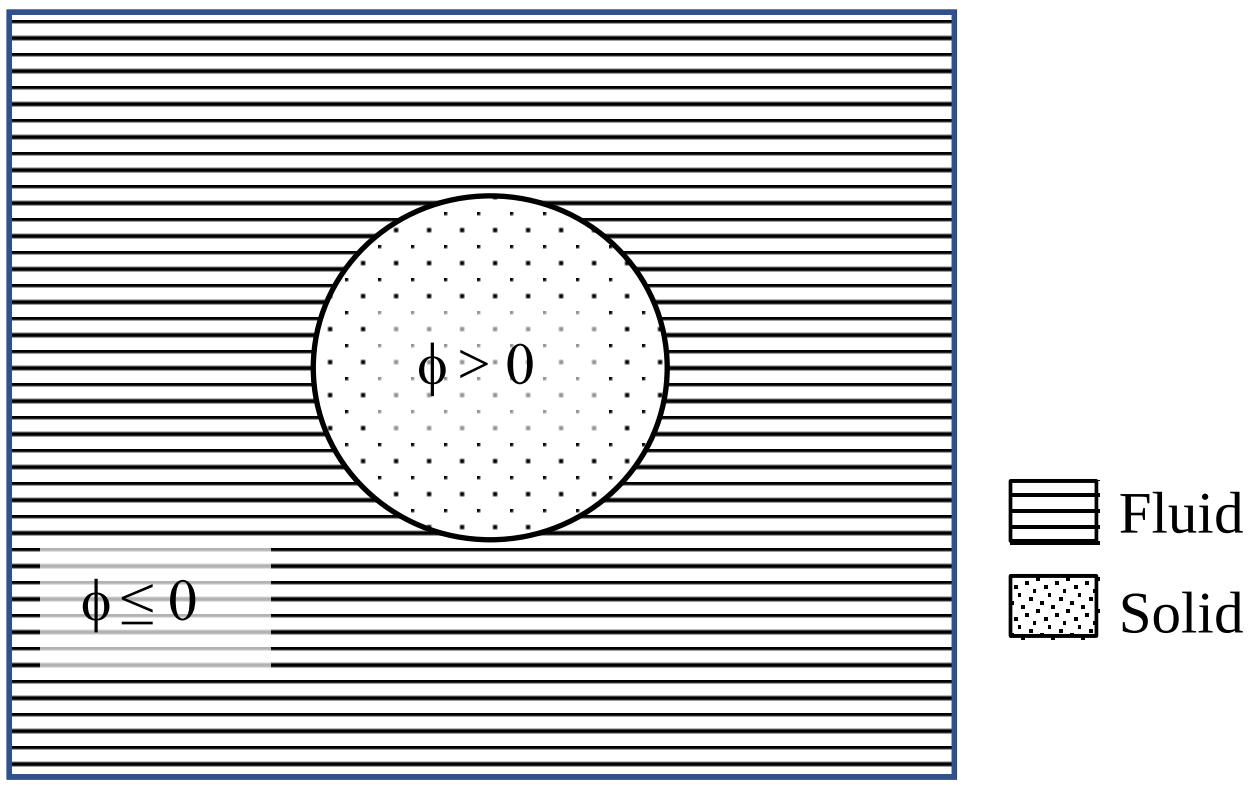

Figure 1: Description of each material subdomain using the indicator function, $\phi(\boldsymbol{x},t)$.

Thus, the density field can be calculated as

$$\rho(\mathbf{x},t) = \rho_f + (\rho_s - \rho_f)H(\phi(\mathbf{x},t)) \tag{8}$$

with

$$H(\phi) = \begin{cases} 0 & if \quad \phi \leq 0 \quad (fluid) \\ 1 & if \quad \phi > 0 \quad (solid) \end{cases} \tag{9}$$

where the subscripts, *f* and *s*, stand for fluid and solid respectively.

The calculation of the rigid motion $\bar{\mathbf{u}}$, is obtained by averaging translation and angular velocities over the solid:

$$\bar{\boldsymbol{u}} = \frac{1}{M}\int_\Omega \rho(\boldsymbol{x},t)H(\phi)\boldsymbol{u}\, d\Omega + \left(\boldsymbol{I}^{-1}\int_\Omega (\boldsymbol{x}-\boldsymbol{x}_G)\times\rho(\boldsymbol{x},t)H(\phi)\boldsymbol{u}\, d\Omega\right)\times(\boldsymbol{x}-\boldsymbol{x}_G) \tag{10}$$

The rigid body moves with an advection velocity that can be chosen to be either the flow velocity $\mathbf{u}$ or the rigid motion $\bar{\mathbf{u}}$ (both values are equivalent just in the limit $\lambda \rightarrow \infty$). However, the choice to follow the solid phase with the velocity $\bar{\mathbf{u}}$ instead of $\mathbf{u}$ is recommended here, since a strictly rigid motion of the solid is guaranteed, independently on numerical errors, a feature which is desirable in practice.

In order to track the solid motion, advection of the solid properties is required. Since density is a material property moving with the flow, its material derivative is zero, and therefore:

$$\frac{D(\rho)}{Dt} = \frac{\partial(\rho)}{\partial t} + \bar{\boldsymbol{u}}\, grad(\rho) \ = 0 \qquad for\ \boldsymbol{x} \in \Omega\ and\ t > 0 \tag{11}$$

In order to solve above system of equations (6)-(11), the following fractional-step algorithm is proposed:

Given a time step $\Delta t$, and being $t^n = n\Delta t$, it is possible to go from $\mathbf{u}^n \approx \mathbf{u}(\cdot, t^n)$ to $\mathbf{u}^{n+1}$ by:

Step 1: Solving Navier-Stokes equation (6)-(7) with $\lambda = 0$ for a time step $\Delta t$; $\tilde{\mathbf{u}}^n$ will be obtained.
Step 2: Computing $\bar{\mathbf{u}}^n$ from (10) after replacing $\mathbf{u}$ by the result $\tilde{\mathbf{u}}^n$ obtained in the previous step.
Step 3: Solving

$$\rho(\mathbf{x},t)\frac{\partial \mathbf{u}}{\partial t} = \lambda H(\phi)\rho(\mathbf{x},t)(\bar{\mathbf{u}} - \mathbf{u}) \tag{12}$$

for a time step $\Delta t$, and choosing $\lambda = 1/\Delta t$, this step returns

$$\mathbf{u}^{n+1} = \begin{cases} \tilde{\mathbf{u}}^n & in\ \mathrm{F(t)} \\ \bar{\mathbf{u}}^n & in\ \mathrm{S(t)} \end{cases} \tag{13}$$

It is important to note here that one of the advantages of considering the penalty formulation is that the penalty equation (12) can be discretized in time in an implicit manner enabling larger penalty coefficients and therefore better accuracy in satisfying the interface boundary conditions. On the contrary, explicit time discretization of (12) would require smaller values for $\lambda$ than $1/\Delta t$ [16,65-66].

Step 4: Advecting the indicator function, $\phi$, with velocity $\bar{\mathbf{u}}$ will allow the determination of the new density field (8), and thus the new position for the solid at $t^{n+1}$

$$\frac{\partial \phi}{\partial t} + \bar{\mathbf{u}}\ grad\ \phi = 0 \tag{14}$$

## 2.3. Modelling the fluid-solid interface via level set technique

One of the main problems found in the fluid-solid interaction is the proper modelling of the interface which separates the solid moving mass from the viscous fluid. In this work a level set technique has been adopted [45,58-61], which will be described next.

The unsteady flow of the interacting fluid and solid is modelled using equations (6)-(11). For the example illustrated in Figure 1, the material property (i.e. density, viscosity), *MP*, in both subdomains may be calculated from $\phi$ as

$$MP(\boldsymbol{x},t) = MP_f + (MP_s - MP_f)H(\phi(\boldsymbol{x},t)) \tag{15}$$

Since *P* is a material property moving with the flow, its material derivative is zero

$$\frac{D(MP)}{Dt} = \frac{\partial(MP)}{\partial t} + \bar{\boldsymbol{u}}\ grad(MP)\ = 0 \tag{16}$$

and considering the dependence of the material properties on the indicator function, (15), this condition may be written as

$$\frac{D\phi}{Dt} = \frac{\partial \phi}{\partial t} + \bar{\boldsymbol{u}}\ grad\phi = 0 \tag{17}$$

which states that the indicator is purely advected by the flow and requires the function *H*(.) to be smooth.

In the case the indicator function, $\phi$, is a linear function of its position, $\mathbf{x}$, its second order spatial derivative is zero and equation (17) is exact, and thus its numerical approximations. Therefore, considering $\phi$ as linear, also benefits the numerical solution of (17) as the front smearing caused by low order numerical schemes or the oscillations induced in the high order case will not appear. The simplest linear function is that with slope unity, i.e. the distance function: $|grad\phi| = 1$.

As regards the definition of *H*(.), equation (9) works well only for small density ratios. For high density ratios, it may result in unwanted instabilities in the pressure field giving rise to highly inaccurate solutions. This problem is related to the numerical solution of a badly conditioned Poisson equation for the pressure [67]. To avoid abrupt changes in the material properties when crossing the interface, function H(.) is interpolated through a constant thickness tube of total width 2δ surrounding the interface [61], where δ is taken of the order of the mesh size. There exist different alternatives for this re-definition of function *H*(.), such as the simple linear interpolation [68]

$$H(\phi) = \begin{cases} 0 \quad for \quad \phi \leq -\delta \\ (\phi + \delta) \quad for \quad -\delta < \phi < \delta \\ 1 \quad for \quad \phi \geq \delta \end{cases} \tag{18}$$

However, extra smoothing can be gained considering other functions. This paper considers the sine function given below [58]

$$H(\phi) = \begin{cases} 0 \quad for \quad \phi \leq -\delta \\ sin\left(\frac{\pi(\phi+\delta)}{4\delta}\right) \quad for \quad -\delta < \phi < \delta \\ 1 \quad for \quad \phi \geq \delta \end{cases} \tag{19}$$

This definition of the interpolation function, based on the distance to the interface, requires keeping the indicator function as a distance function. In effect, equation (17) states that the indicator is advected by the fluid velocity: as the fluid velocity is not uniform in the domain, the initial distance function will be distorted as time progresses and, after some time, it will not be a distance function any longer.

This issue can be addressed by using the fluid flow velocity as proposed in equation (17) to advect the indicator. Once the indicator is advected, it must be corrected in order to comply with the $|grad\phi| = 1$ condition. This can be achieved by solving at any time *t*, the following problem to steady state [61]:

$$\frac{\partial\phi(\hat{\tau})}{\partial\hat{\tau}} + \mathcal{S}(\phi(t))|grad\ \phi(\hat{\tau})| = \mathcal{S}(\phi(t)) \tag{20}$$

with initial conditions

$$\phi(\boldsymbol{x}, \hat{\tau} = 0) = \phi(\boldsymbol{x}, t) \tag{21}$$

being $\mathcal{S}(.)$ the sign function and $\hat{\tau}$ a fictitious time.

Clearly, the steady-state solution of this problem is compliant with the condition $|grad\phi| = 1$ and the zero level set of $\phi(\hat{\tau} \to \infty)$ matches that of $\phi(t)$.

Equation (20) may be written also as

$$\frac{\partial\phi(\hat{\tau})}{\partial\hat{\tau}} + \mathcal{S}(\phi^n)\frac{grad\ \phi(\hat{\tau})}{|grad\ \phi(\hat{\tau})|}grad\ \phi(\hat{\tau}) = \mathcal{S}(\phi^n) \tag{22}$$

which is an advection problem with velocity

$$\boldsymbol{v} = \mathcal{S}(\phi^n)\frac{grad\ \phi(\hat{\tau})}{|grad\ \phi(\hat{\tau})|} \tag{23}$$

This equation for the velocity indicates that the problem characteristics initiate at the interface position and travel with velocity ±1. Therefore, reconstruction of the indicator function as a distance function initiates at the interface position and progresses along its outward normal direction. Thus, the critical zone, surrounding the interface position, is reconstructed in the first iterations of the solution of problem (22).

This idea may be easily extended to MS solids in a multiphase flow of NF immiscible fluids by incorporating NF+MS-1 indicator functions, as it is shown in Table 1.

| | **Fluid 1** | **Fluid 2** | **…** | **Fluid NF** | **Solid 1** | **…** | **Solid MS** |
|---|---|---|---|---|---|---|---|
| $\boldsymbol{\phi_1}$ | $\phi_1 \leq 0$ | $\phi_1 > 0$ | … | $\phi_1 > 0$ | $\phi_1 > 0$ | … | $\phi_1 > 0$ |
| $\boldsymbol{\phi_2}$ | $\phi_2 \leq 0$ | $\phi_2 \leq 0$ | … | $\phi_2 > 0$ | $\phi_2 > 0$ | … | $\phi_2 > 0$ |
| $\boldsymbol{\phi_3}$ | $\phi_3 \leq 0$ | $\phi_3 \leq 0$ | … | $\phi_3 > 0$ | $\phi_3 > 0$ | … | $\phi_3 > 0$ |
| **…** | … | … | … | … | … | … | … |
| $\boldsymbol{\phi_{NF}}$ | $\phi_{NF} \leq 0$ | $\phi_{NF} \leq 0$ | … | $\phi_{NF} \leq 0$ | $\phi_{NF} > 0$ | … | $\phi_{NF} > 0$ |
| $\boldsymbol{\phi_{NF+1}}$ | $\phi_{NF+1} \leq 0$ | $\phi_{NF+1} \leq 0$ | … | $\phi_{NF+1} \leq 0$ | $\phi_{NF+1} \leq 0$ | … | $\phi_{NF+1} > 0$ |
| $\boldsymbol{\phi_{NF+2}}$ | $\phi_{NF+2} \leq 0$ | $\phi_{NF+2} \leq 0$ | … | $\phi_{NF+2} \leq 0$ | $\phi_{NF+2} \leq 0$ | … | $\phi_{NF+2} > 0$ |
| **…** | … | … | … | … | … | … | … |
| $\boldsymbol{\phi_{NF+MS-1}}$ | $\phi_{NF+MS-1} \leq 0$ | $\phi_{NF+MS-1} \leq 0$ | … | $\phi_{NF+MS-1} \leq 0$ | $\phi_{NF+MS-1} \leq 0$ | | $\phi_{NF+MS-1} > 0$ |

Table 1: Assignment of each material subdomain using NF+MS-1 indicator functions.

However, for a system of MS rigid bodies interacting in an incompressible multi-fluid flow, a special method to handle collisions should be designed, which does not fall within the scope of this work. Thus, in the following, only one solid immersed in a viscous fluid will be considered.

## 3. Numerical model

The numerical method used to solve the Navier–Stokes equations (6)-(7) and the advection (17) and correction (22) of the indicator function is the two-step Taylor-Galerkin method. This method is described in [49], where the interested reader can find the details of the derivation.

### 3.1. Step 1: Solving Navier-Stokes equations with $\lambda = 0$

In order to solve Navier-Stokes equations (6)-(7) with $\lambda = 0$ for a time step $\Delta t$, the fractional-step procedure proposed by Chorin [44] is followed.

The velocity is decomposed into two parts

$$\Delta \mathbf{u}^{\mathrm{n}} = \Delta \mathbf{u}^{*,\mathrm{n}} + \Delta \mathbf{u}^{**,\mathrm{n}} \tag{24}$$

such that

$$\rho \frac{\Delta \mathbf{u}^{*,n}}{\Delta t} + \rho \, div(\mathbf{u}^n \otimes \mathbf{u}^n) - div \, \boldsymbol{\tau}^n - \rho \mathbf{g}^n = 0 \tag{25}$$

$$\rho \frac{\Delta \mathbf{u}^{**,n}}{\Delta t} + grad \, p^{n+1} = 0 \tag{26}$$

along with the continuity equation

$$div \, \mathbf{u}^{n+1} = 0 \tag{27}$$

being $\boldsymbol{u}^{*,n}$ and $\boldsymbol{u}^{**,n}$ the intermediate velocities resulting from (25) and (26) respectively.

Concerning the spatial discretization, 2D linear triangles have been chosen because of their numerical efficiency and excellent behaviour in the solution of problems involving strong discontinuities [47-48,50-52]. However, Babuska-Brezzi condition [69-70] does not allow the use of the same order of interpolation for both velocity and pressure unless a special stabilization technique is used. As it was shown in [49,71], the proposed algorithm provides with the required stabilization.

Therefore, above resultant equations from the time discretization, are discretized in space as follows:

***Step I: Fractional velocity discretization***. The fractional linear momentum equation

$$\rho\frac{\Delta\mathbf{u}^{*,n}}{\Delta t}+\rho\ div(\mathbf{u}^n\otimes\mathbf{u}^n)-div\ \boldsymbol{\tau}^n-\rho\mathbf{g}^n=0 \tag{28}$$

is discretized in space using the two-step Taylor-Galerkin algorithm:

$$(u_i)^{*,n+1/2}=(u_i)^n-\frac{\Delta t}{2}\left(\frac{\partial(u_ju_i)^n}{\partial x_j}-g_i^n\right) \tag{29}$$

$$\boldsymbol{M}\Delta u_i{}^{*,n}=\Delta t\left\{\int_\Omega\left[\frac{\partial\boldsymbol{N}^T}{\partial x_j}(u_ju_i)^{*,n+1/2}-\frac{1}{\rho}\frac{\partial\boldsymbol{N}^T}{\partial x_j}(\tau_{ij})^n\right]d\Omega+\right.$$
$$\left.+\int_\Omega\boldsymbol{N}^T(g_i)^{*,n+1/2}\,d\Omega-\int_\Gamma\boldsymbol{N}^T\left((u_ju_i)^{*,n+1/2}-\frac{1}{\rho}(\tau_{ij})^n\right)\cdot n_jd\Gamma\right\} \tag{30}$$

where **N** is the shape function which interpolates the solution between the discrete values for velocity and pressure at the mesh nodes.

***Step II: Continuity equation discretization.*** Recalling the continuity equation at $t^{n+1}$ given in (27) and using the incremental momentum split (26), discretization of the continuity equation can be written as

$$div\ \boldsymbol{u}^{*,n}-\Delta t\ div\left(\frac{1}{\rho}grad\ p^{n+1}\right)=0 \tag{31}$$

where $\boldsymbol{u}^{*,n}=\boldsymbol{u}^n+\Delta\boldsymbol{u}^{*,n}$.

Applying the standard Galerkin discretization, spatial discretization of above equation is

$$\int_\Omega\frac{1}{\rho}\ (grad\,\boldsymbol{N})^Tgrad\,\boldsymbol{N}\ d\Omega\ \ \Delta\bar{p}^n=-\frac{1}{\Delta t}\int_\Omega\boldsymbol{N}^Tdiv\,\boldsymbol{u}^{*,n}\,d\Omega-$$
$$-\int_\Omega\frac{1}{\rho}(grad\,\boldsymbol{N})^Tgrad\,p^n\,d\Omega+\int_{\Gamma-\Gamma_p}\frac{1}{\rho}\boldsymbol{N}^Tgrad\,p^{n+1}\cdot\boldsymbol{n}\,d\Gamma \tag{32}$$

where pressure is prescribed at $\Gamma_p$.

In order to compute the boundary integral, equation (31) is projected along the normal direction,

$$\frac{1}{\rho}grad\ p^{n+1}\cdot\boldsymbol{n}=-\frac{1}{\Delta t}[\boldsymbol{u}^{n+1}-(\boldsymbol{u}^n+\Delta\boldsymbol{u}^{*,n})]\cdot\boldsymbol{n} \tag{33}$$

resulting in

$$\int_{\Gamma-\Gamma_p}\frac{1}{\rho}\boldsymbol{N}^Tgrad\ p^{n+1}\cdot\boldsymbol{n}d\Gamma=-\frac{1}{\Delta t}\int_{\Gamma-\Gamma_p}\boldsymbol{N}^T[\boldsymbol{u}^{n+1}-(\boldsymbol{u}^n+\Delta\boldsymbol{u}^{*,n})]\cdot\boldsymbol{n}d\Gamma \tag{34}$$

Therefore, the system of equations to be solved to account for the pressure increment is

$$\int_\Omega\frac{1}{\rho}(grad\,\boldsymbol{N})^Tgrad\,\boldsymbol{N}\ d\Omega\ \ \Delta\bar{p}^n=-\frac{1}{\Delta t}\int_\Omega\boldsymbol{N}^Tdiv\ \boldsymbol{u}^{*,n}d\Omega-$$
$$-\int_\Omega\frac{1}{\rho}(grad\,\boldsymbol{N})^Tgrad\,p^n\,d\Omega+\int_{\Gamma-\Gamma_p}\boldsymbol{N}^T[\boldsymbol{u}^{n+1}-(\boldsymbol{u}^n+\Delta\boldsymbol{u}^{*,n})]\cdot\boldsymbol{n}\,d\Gamma \tag{35}$$

***Step III: Velocity correction***. Once pressure has been computed in the previous step, the velocity increment, $\boldsymbol{\Delta u}^*$, must be corrected by discretizing in space equation (26)

$$\int_\Omega \frac{\rho}{\Delta t}\boldsymbol{N}^T\boldsymbol{N}d\Omega\,\Delta\boldsymbol{u}^{**,n} + \int_\Omega \boldsymbol{N}^T grad\, p^{n+1} d\Omega = 0 \quad (36)$$

Solving above system of equations $\Delta\boldsymbol{u}^{**,n}$ is obtained, which is added to $\Delta\boldsymbol{u}^{*,n}$, resulting in the total velocity increment $\Delta\boldsymbol{u}^n$:

$$\Delta\boldsymbol{u}^n = \Delta\boldsymbol{u}^{*,n} + \Delta\boldsymbol{u}^{**,n} \quad (37)$$

and thus,

$$\tilde{\boldsymbol{u}}^{\boldsymbol{n}} = \boldsymbol{u}^n + \Delta\boldsymbol{u}^n \quad (38)$$

The proposed algorithm can be extended to the 3-dimensional problem by just using linear tetrahedra elements.

The maximum allowed time step in the solution of Navier-Stokes equations is [49]:

$$C \le \beta\sqrt{\frac{1}{Pe^2} + \alpha} - \frac{1}{Pe} \quad (39)$$

where:

· C is the Courant number: C=|**u**|/($h_e$/Δ$t$)
· Pe is the Peclet number: Pe=|**u**|$h_e$/(2$\mu$/$\rho$)
· $h_e$ is the element size (minimum height of each triangle has been considered here).
· $\alpha$ =1 when using the lumped mass matrix and $\alpha$ =1/3 when using the consistent mass matrix in the solution of steps I and III.
· $\beta$ is a safety coefficient, typically 0.85-0.9

Thus, the global time step limit is calculated as the minimum time step allowed in the mesh. Equation (39) incorporates the effects of viscosity via Peclet number and, in order to account for non-linearities, a safety factor $\beta$ is considered.

### 3.2. Step 2: Calculating the rigid motion

According to equation (10), calculation of the rigid motion $\bar{\mathbf{u}}$, is obtained by averaging translation and angular velocities over the solid, S, where the density field is given by (8).

Applying the Finite Element Method basic theory, the nodal velocity in the solid, $\bar{\boldsymbol{u}}^{\boldsymbol{n}}$, is computed as:

$$\bar{\boldsymbol{u}}^{\boldsymbol{n}} = \frac{\rho_S}{M}\sum_e[\tilde{\boldsymbol{u}}_e^n]\,\Omega_e + \left[\frac{\rho_S}{\boldsymbol{I}}\sum_e[(\boldsymbol{x}_e - \boldsymbol{x}_G)\times\tilde{\boldsymbol{u}}_e^n]\Omega_e\right]\times(\boldsymbol{x}_i - \boldsymbol{x}_G) \quad (40)$$

where the sub-index $e$ stands for "element" and $G$ for gravity center of the element; $M$ is the mass of the solid, $\Omega_e$ is the element volume (or surface), and **I** is the inertia matrix; $\tilde{\boldsymbol{u}}_{\boldsymbol{e}}$ is the velocity computed in the previous step calculated at the element level and $\rho_S$ is the density of the solid.

### 3.3. Step 3: Obtaining the final velocity

In order to get the final velocity at $t^{n+1}$, equation (12) is solved. Applying the standard Galerkin discretization:

$$\int_\Omega \mathbf{N}^T\rho_S\mathbf{N}\,d\Omega\,\Delta\mathbf{u}^n = \lambda\Delta t\int_\Omega \mathbf{N}^T\rho_S(\bar{\mathbf{u}}^n - \mathbf{u}^n)\mathbf{N}\,d\Omega \quad (41)$$

and $\mathbf{u}^{n+1}$ is obtained in the entire domain, such that $\boldsymbol{u}^{n+1} = \boldsymbol{u}^{n} + \Delta \boldsymbol{u}^{n}$.

As mentioned in Section 2.2, the time step restriction for the proposed explicit time discretization of (41) prevent from using values of λ larger than $1/\Delta t$ [16,65-66]. This issue will be further discussed in Section 4.3.1.

### 3.4. Step 4: Advecting the solid properties

The objective of the indicator function is assigning fluid/solid properties at each point within the domain. The algorithm used for tracking the interface is based on (i) advection of the indicator function using the computed velocity field (17) and (ii) correction of the indicator function in order to keep it as distance function by obtaining the stationary solution of (21)-(22).

***Advection***. Advection equation (17) is discretized using the two-step Taylor-Galerkin algorithm:

$$\phi^{n+1/2} = \phi^{n} - \frac{\Delta t}{2}(\bar{\boldsymbol{u}}\ grad\ \phi^{n})$$
$$\boldsymbol{M}\Delta\bar{\phi}^{n} = \Delta t\left\{\int_{\Omega}(grad^{T}(\bar{\boldsymbol{u}}\boldsymbol{N}))^{T}\phi^{n+1/2}d\Omega - \int_{\Gamma}\boldsymbol{N}^{T}\bar{\boldsymbol{u}}\phi^{n+1/2}\boldsymbol{n}d\Gamma\right\} \tag{42}$$

***Correction***. Once the indicator function has been advected, it is maintained as a distance function by obtaining the stationary solution of (21)-(22). Equation (22) is indeed an advection equation with velocity (23) and a source term, $\mathcal{S}(\phi^{n})$.

Therefore, equation (22) may be written as

$$\frac{\partial\phi(\hat{\tau})}{\partial\hat{\tau}} + \mathbf{v}\ grad\phi(\hat{\tau}) = \mathcal{S}(\phi^{n}) \tag{43}$$

and can be solved using the two-step Taylor-Galerkin algorithm:

$$\phi^{n+1/2} = \phi^{n} + \frac{\Delta\hat{\tau}}{2}[\mathcal{S}(\phi^{n}) - \boldsymbol{v}\ grad\ \phi^{n}]$$
$$\boldsymbol{M}\Delta\bar{\phi}^{n} = \Delta\hat{\tau}\left\{\int_{\Omega}\boldsymbol{N}\mathcal{S}(\phi^{n+\frac{1}{2}})d\Omega + \int_{\Omega}(grad^{T}(\boldsymbol{v}\boldsymbol{N}))^{T}\phi^{n+1/2}d\Omega - \int_{\Gamma}\boldsymbol{N}^{T}\boldsymbol{v}\phi^{n+1/2}\boldsymbol{n}d\Gamma\right\} \tag{44}$$

Once the indicator function has been advected and corrected, the material properties, *MP*, are interpolated (15) using the sine function given in (19).

In order to accelerate convergence of equation (43) to the steady state, the algorithm uses the lumped mass matrix along with an optimum time step for each mesh element [49]:

$$\Delta\hat{\tau} = \beta\frac{h}{|\mathbf{v}|} = \beta h \tag{45}$$

being $\beta$ a safety coefficient lower than 1.

Nevertheless, the correction phase implies a significative computational cost. In order to reduce this computational effort, both phases, advection and correction, are limited to a thin region about the zero level set of the indicator function. Following this procedure, the number of iterations is drastically reduced since condition $|\mathrm{grad}\phi| = 1$ must be fulfilled only within this thin region. The size of the mentioned region is about 3δ - 4δ centred at the interface.

## 4. Verification results

In this section, some numerical examples have been selected in order to validate the method since either they have been studied experimentally or they have already been solved using other numerical solutions.

The examples considered next have been chosen in order to illustrate the main advantages of the proposed model:

- The method is capable to reproduce the experimental results of the flow dynamics around a fixed body for a wide range of Reynolds numbers.
- The model is able to track the fluid-solid interface without distortion for a very high number of iterations, maintaining the initial volume and shape of the solid.
- The proposed model is able to predict the dynamics of an immersed body with a high level of accuracy.
- The computational cost to get accurate results is very low.

### 4.1 Flow around a circular cylinder

This problem has been widely studied by many researchers in the past [72-84] giving rise to a variety of experimental results, empirical formulae and advanced numerical methods. Thus, detailed analysis of the flow around a cylinder is of high interest allowing validation of models against experimental and computational results, and becoming the starting point for further applications to the study of more complex geometries such as planes, ships or submarines.

In order to assess the performance of the proposed model when dealing with the flow around a fixed rigid body, the classic problem of the flow around a circular cylinder placed on a uniform flow will be analysed here. This problem consists of a solid circular cylinder of diameter D = 1 m and density $\rho_s$ = 3000 kg/m$^3$ fixed and immersed in a viscous fluid flow which is moving with uniform horizontal velocity, as sketched in Figure 2.

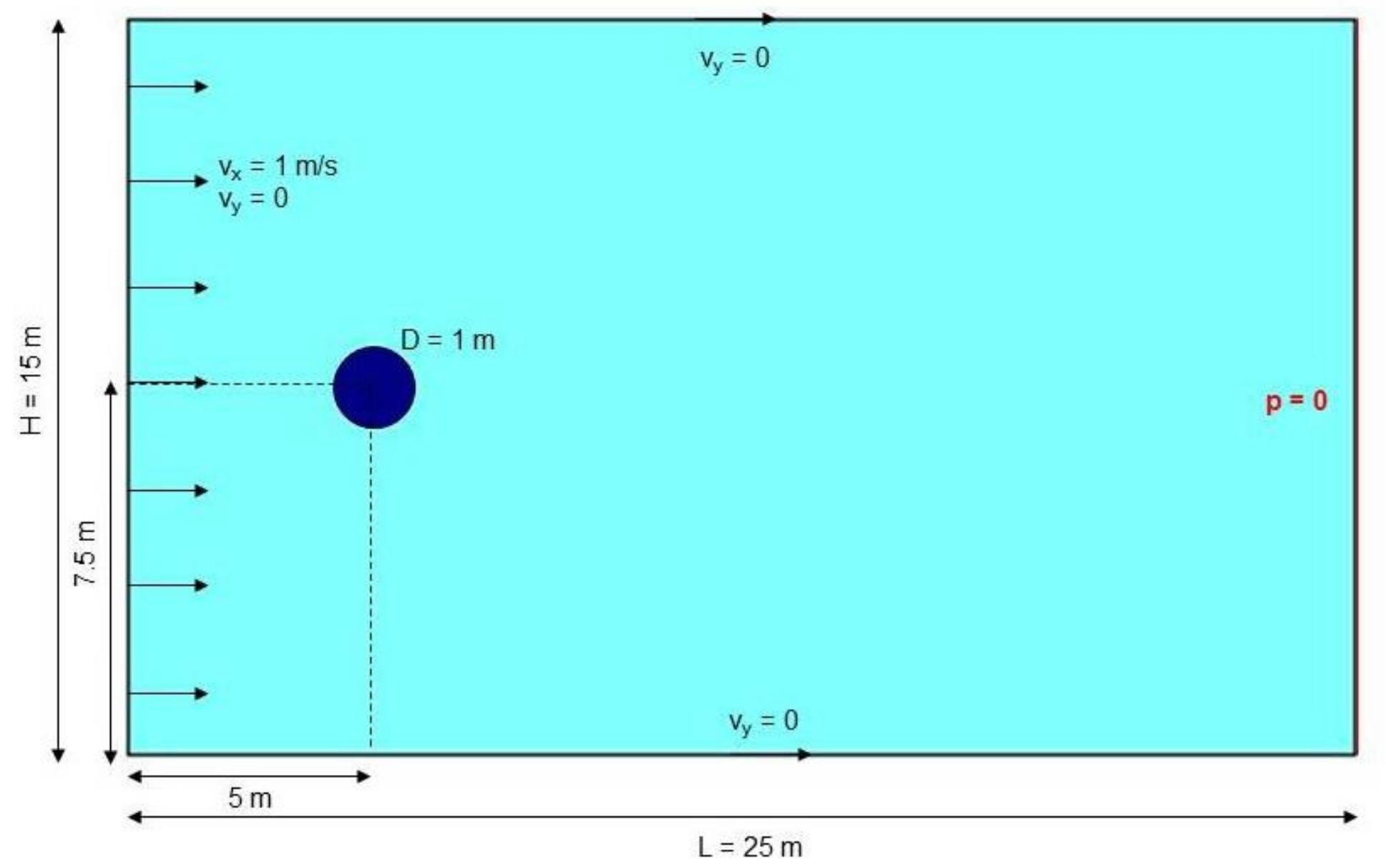

Figure 2: Flow around a circular cylinder: problem layout and boundary conditions

A sensitivity analysis of the proposed method with respect to the mesh size and the time step is considered next in order to properly select the optimal mesh size and time step for the analysed problem.

The theoretical inviscid pressure coefficient, $c_p$, on a circular cylinder is [81]:

$$c_p = \frac{p - p_\infty}{\frac{1}{2}\rho v_\infty^2} = 1 - 4sin^2\theta \quad (46)$$

In order to analyse the mesh dependence on the solution, the inviscid pressure coefficient on the cylinder has been computed for different meshes. The number of mesh nodes has been gradually increased and the accuracy of the solution for different choices of the mesh size has been studied. The sensitivity analysis of the numerical solution as a function of the number of nodes is given in Figure 3. The values considered in this study are within the range of 1387–2432 nodes. It can be observed in Figure 3 that for the case of 2432 nodes the solution preserves its accuracy, being in good agreement with the analytical solution before detachment of the flow occurs (120º < θ < 130º). However, as the mesh size increases the solution loses its accuracy until it becomes highly distorted for a value of 1387 nodes. The time steps used for the calculations have been chosen according to (39) so that it is optimal for each considered case.

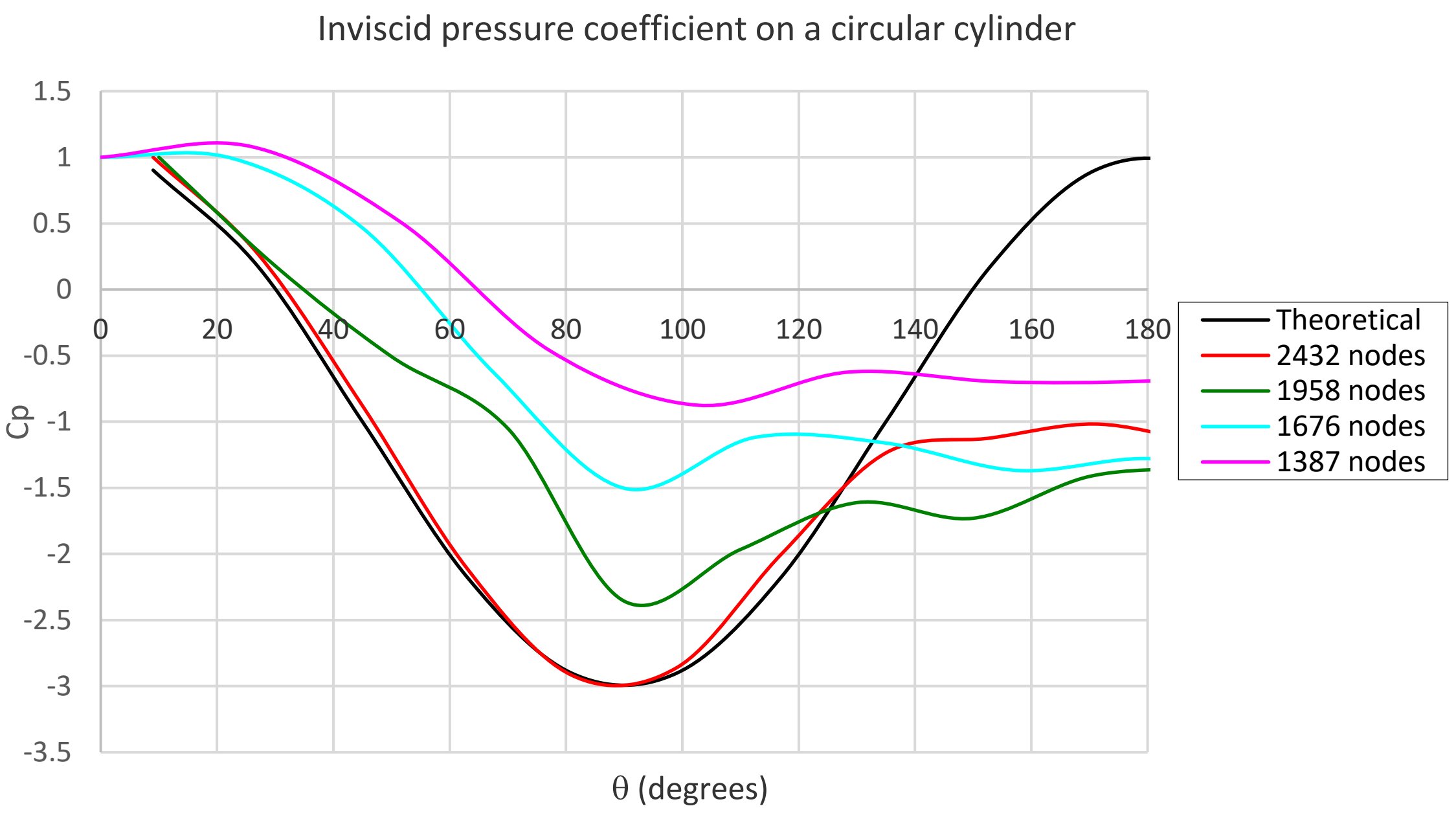

Figure 3: Flow around a circular cylinder: sensitivity of the solution to the mesh size

To accomplish a sensitivity analysis of the proposed method with respect to the time step, Δt, a similar analysis as above has been carried out. Therefore, the same problem of the inviscid pressure coefficient on the cylinder has been solved considering a fixed mesh of 2432 nodes. The value of parameter Δt has been gradually increased and the accuracy of the solution for different values of Δt has been studied. The sensitivity analysis of the numerical solution as a function of the time step, Δt, is given in Figure 4. It can be observed that when Δt is equal to its optimal value given by equation (39) (i.e. Δt = 3.3 $10^{-2}$ s), the accuracy of the numerical solution gets maximal. As the value of the Δt is either increased or decreased with respect its optimal value the solution loses its accuracy. For larger values than Δt = 8 $10^{-2}$ s the numerical solution becomes unstable.

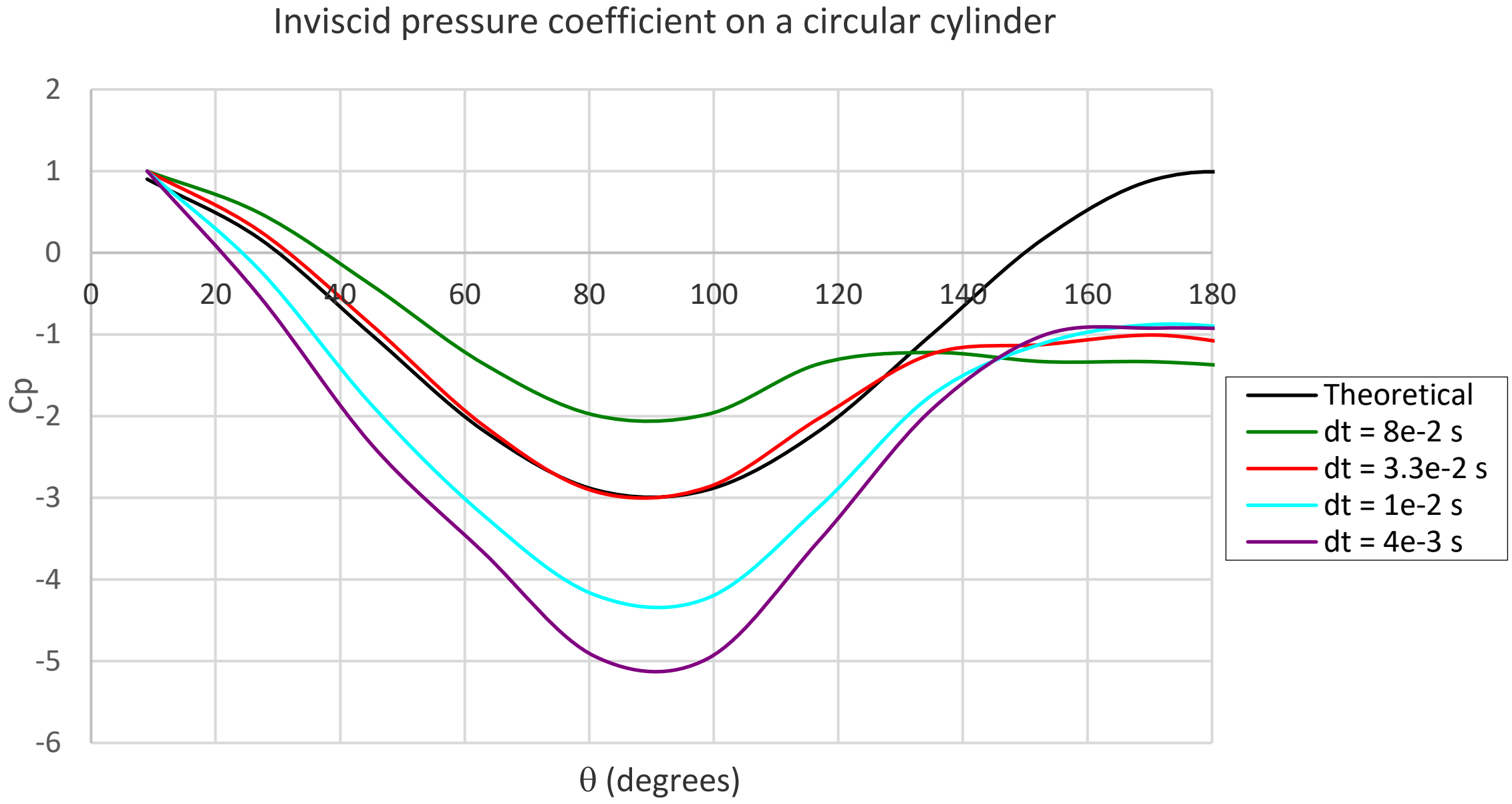

Figure 4: Flow around a circular cylinder: sensitivity of the solution to the time step

Along with the sensitivity analysis, an estimation of the computational cost has been performed. In order to do that, the calculation time for the cases presented in Figures 3 and 4 has been computed. The CPU time for 212 s of simulation using a 2.5 GHz CPU and 4Gb RAM machine are shown in Tables 2-4: for a fixed Δt (Table 2), for a fixed mesh (Table 3), and using the optimal time step given by equation (39) (Table 3). CPU times range between 13 s and 1000 s, while for an optimal mesh size and time step, the CPU time is around 131 s. These results show the efficiency and low computational cost of the proposed method.

| **Fixed Δt (s)** | **Elements** | **Nodes** | **Simulation time (s)** | **CPU time (s)** |
|---|---|---|---|---|
| 3.3e-2 | 4712 | 2432 | 212 | 130.70 |
| 3.3e-2 | 3764 | 1958 | 212 | 103.18 |
| 3.3e-2 | 3200 | 1676 | 212 | 90.25 |
| 3.3e-2 | 2622 | 1387 | 212 | 71.98 |

Table 2: Flow around a circular cylinder: CPU times for a fixed time step

| **Δt (s)** | **Fixed Elements** | **Fixed Nodes** | **Simulation time (s)** | **CPU time (s)** |
|---|---|---|---|---|
| 8e-2 | 4712 | 2432 | 212 | 39.89 |
| 3.3e-2 | 4712 | 2432 | 212 | 130.70 |
| 1e-2 | 4712 | 2432 | 212 | 379.07 |
| 4e-3 | 4712 | 2432 | 212 | 1000.54 |

Table 3: Flow around a circular cylinder: CPU times for a fixed mesh

| **Optimal Δt (s)** | **Elements** | **Nodes** | **Simulation time (s)** | **CPU time (s)** |
|---|---|---|---|---|
| 3.3e-2 | 4712 | 2432 | 212 | 130.70 |
| 4.5e-2 | 3764 | 1958 | 212 | 80.98 |
| 7e-2 | 3200 | 1676 | 212 | 46.99 |
| 1e-1 | 2622 | 1387 | 212 | 13.11 |

Table 4: Flow around a circular cylinder: CPU times for different mesh sizes using the optimal time step

Following the sensitivity analysis results, a non-structured mesh of 4712 linear triangles (2432 nodes) is used for the computation (Figure 5). A horizontal uniform velocity is prescribed in the left boundary of the domain, $v_x$ = 1 m/s and $v_y$ = 0, while $v_y$ = 0 is prescribed in the top and bottom boundaries. Pressure is only prescribed in the right boundary and set equal 0 (see boundary conditions in Figure 2).

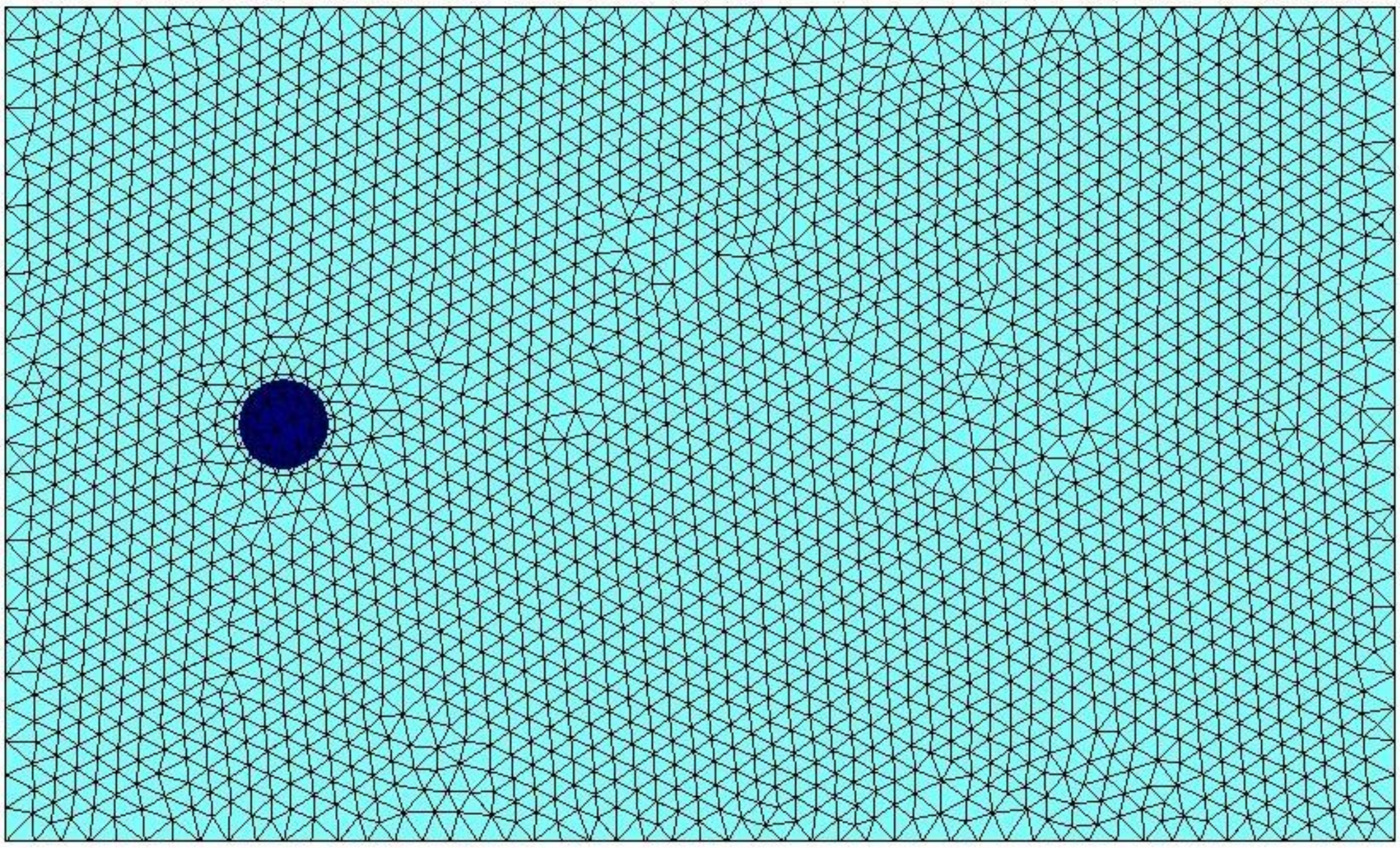

Figure 5: Flow around a circular cylinder: computational mesh of 4712 linear triangles (2432 nodes)

The parameters used in the computation are taken as follows: length and height of the fluid domain are L = 25 m and H = 15 m respectively (Figure 2); the fluid is assumed to be of newtonian type with density $\rho_f$ = 1000 kg/m$^3$. The center of the cylinder is placed at a distance of 5 m from the left side of the domain and 7.5 m high (Figure 2). The solid velocity is set to 0 throughout the whole simulation. The time-step used for the calculation is $\Delta t = 5\ 10^{-2}$ s.

In order to analyse the dynamics of the flow around the cylinder as a function of the Reynolds number, $Re = \frac{\rho v_\infty D}{\mu}$, different values of viscosity $\mu$ will be used in the computations.

For Re = 10: A non-oscillatory solution is reached. Behind the cylinder the boundary layer begins to be detached and two eddies start to be formed spinning in opposite directions; behind these two eddies the streamlines get close again parallel and symmetric. These results are the expected ones for Reynolds numbers in the range 1 < Re < 30 according to the literature [78-79]. Figures 6 and 7 show the velocity vectors and pressure values in the whole domain for t = 212 s.

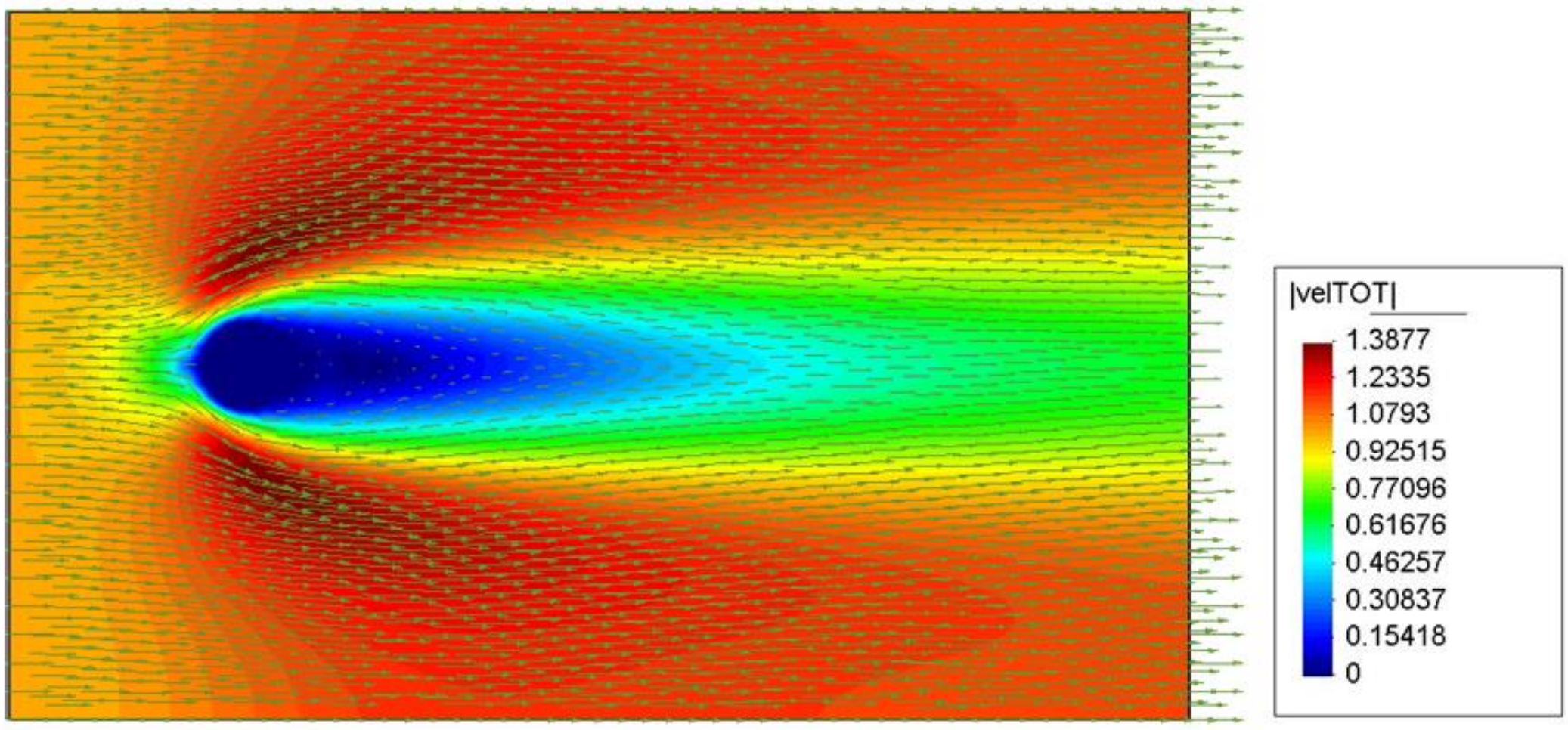

Figure 6: Flow around a circular cylinder: velocity vectors in the whole domain for Re = 10 (t = 212 s)

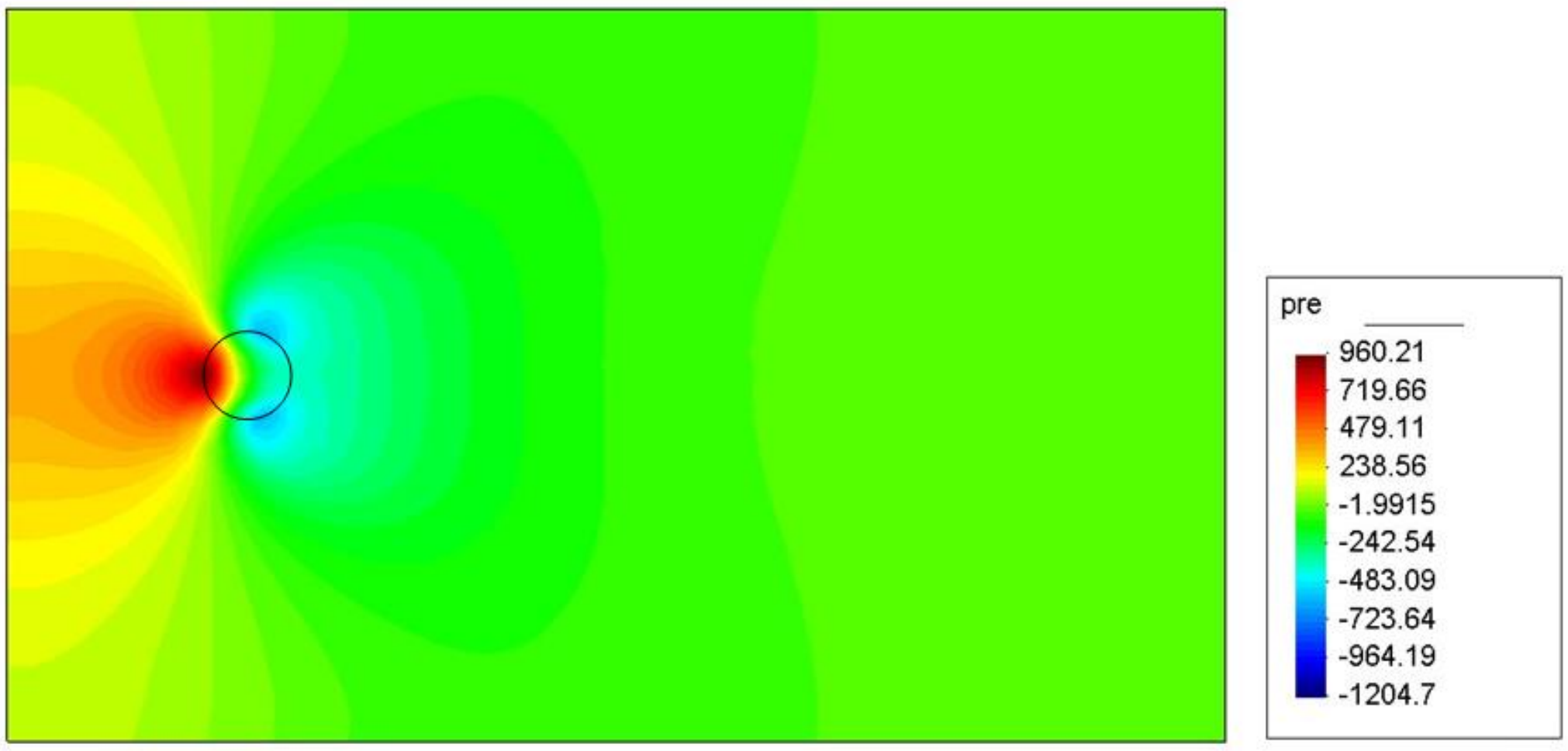

Figure 7: Flow around a circular cylinder: pressure values in the whole domain for Re = 10 (t = 212 s)

For Re = 30: Behind the cylinder, slight periodic oscillations in the flow begin to be detected. These results are expected for Reynolds numbers within the range of 30 < Re < 90 according to [78-79]. Figures 8 and 9 show the velocity field and pressure values in the whole domain for t = 212 s. In Figure 10 the detailed velocity field in the rear of the cylinder is depicted.

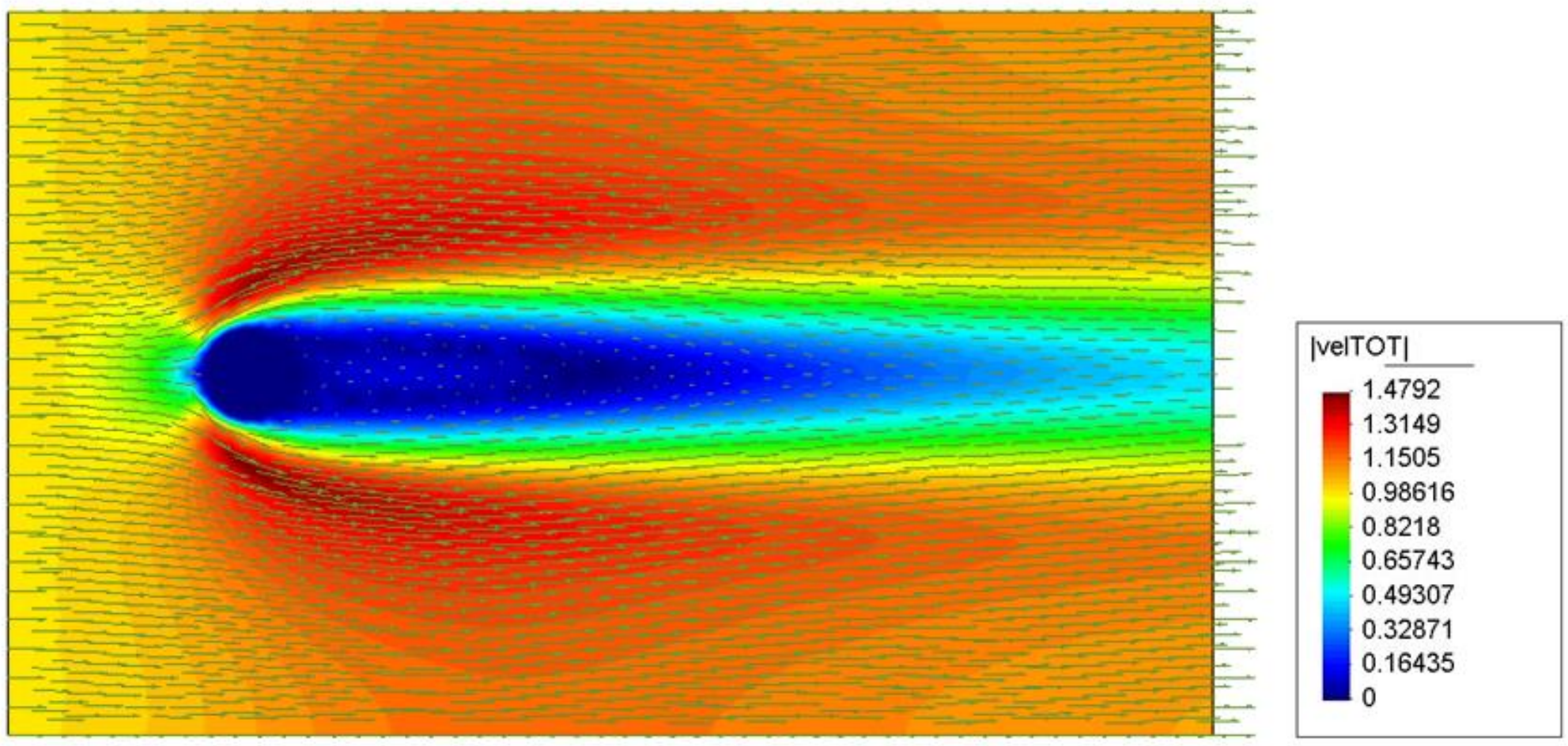

Figure 8: Flow around a circular cylinder: velocity vectors in the whole domain for Re = 30 (t = 212 s)

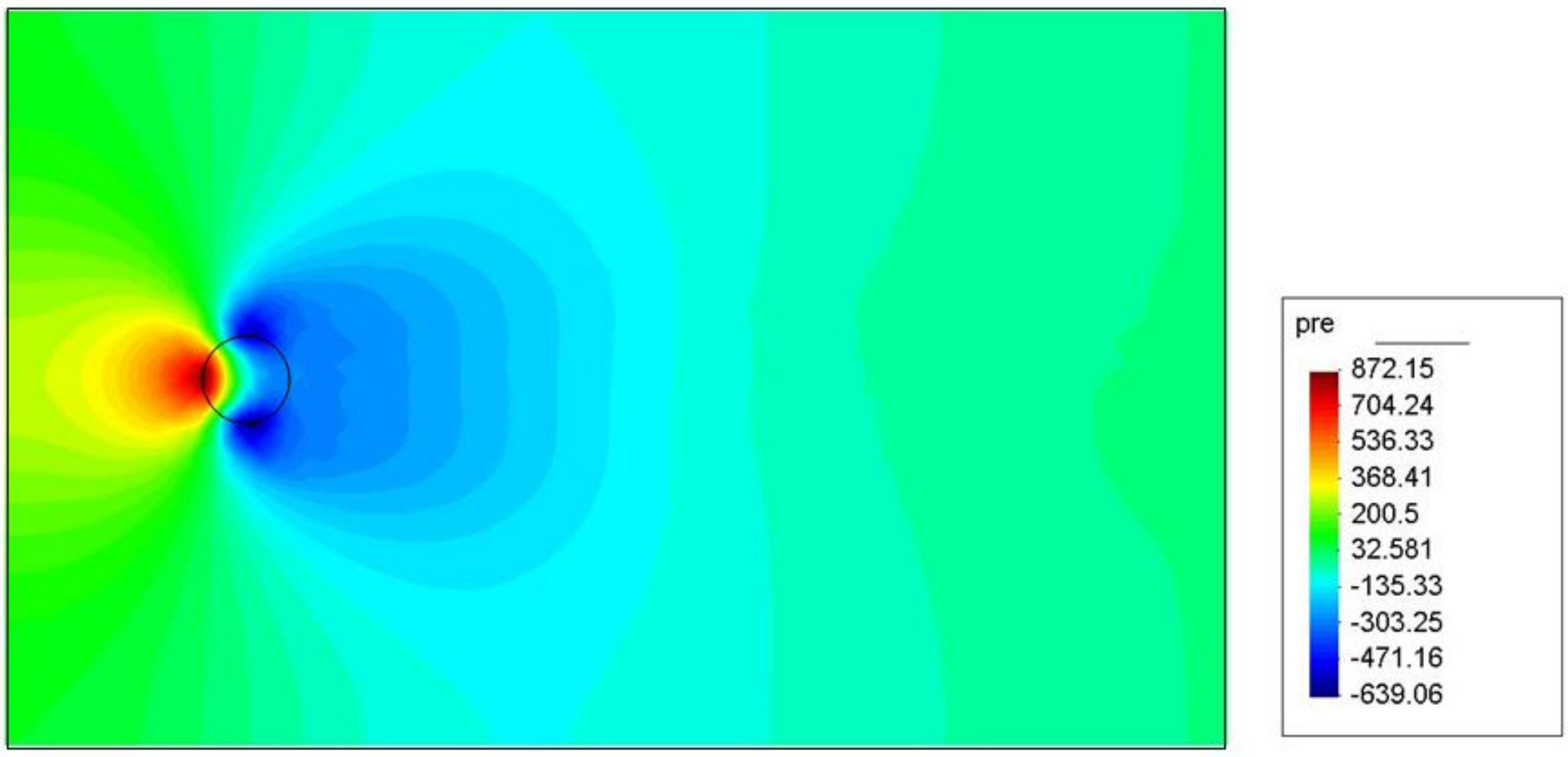

Figure 9: Flow around a circular cylinder: pressure values in the whole domain for Re = 30 (t = 212 s)

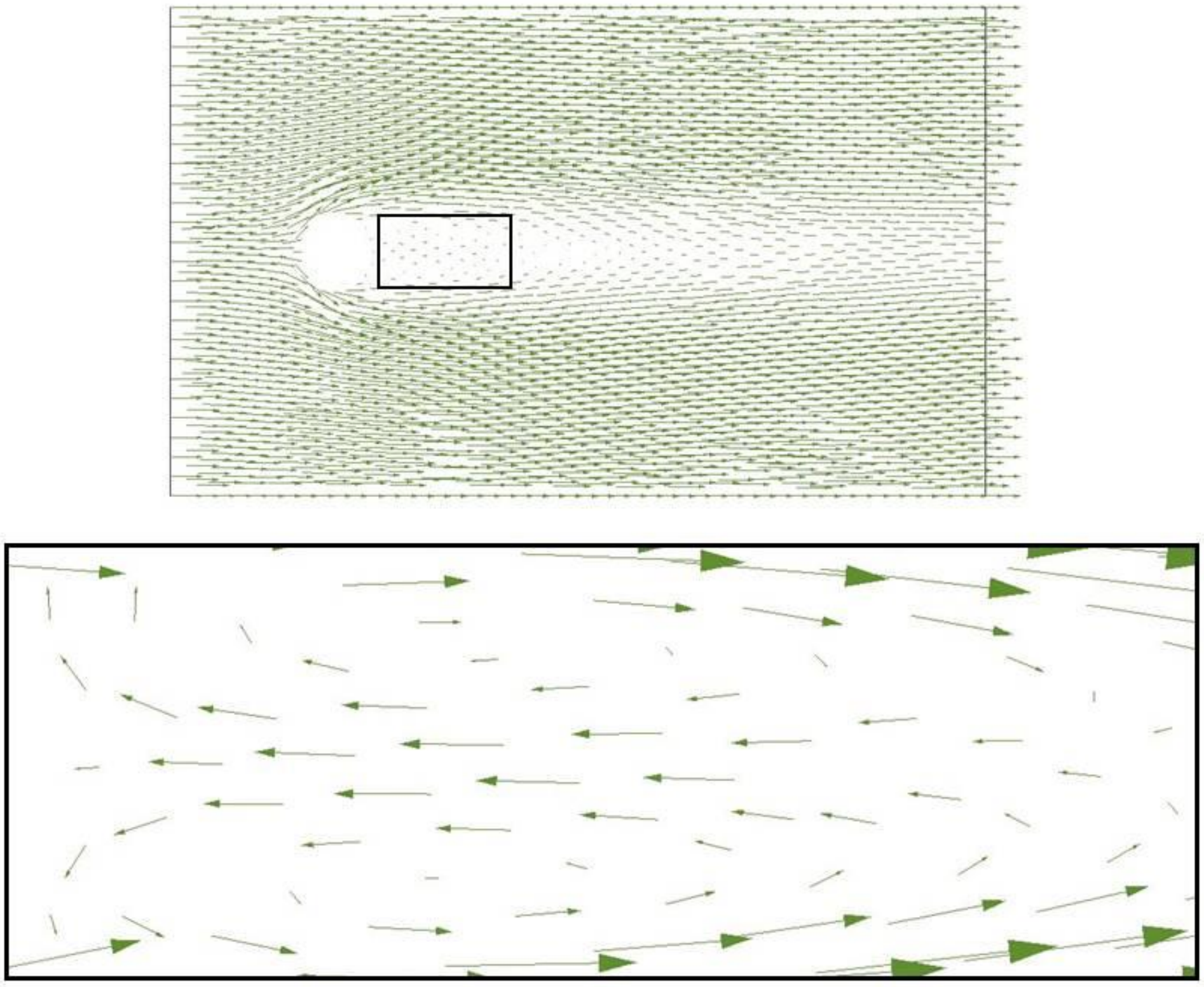

Figure 10: Flow around a circular cylinder: velocity vectors in the rear of the cylinder for Re = 30 (t = 212 s)

For Re = 90: Two eddies begin to be detached alternatively at both sides of the middle axis in the rear wake, being continuously regenerated with a period of 5.5 s. Figures 11 and 12 show the velocity field and pressure values in the whole domain for t = 212 s.

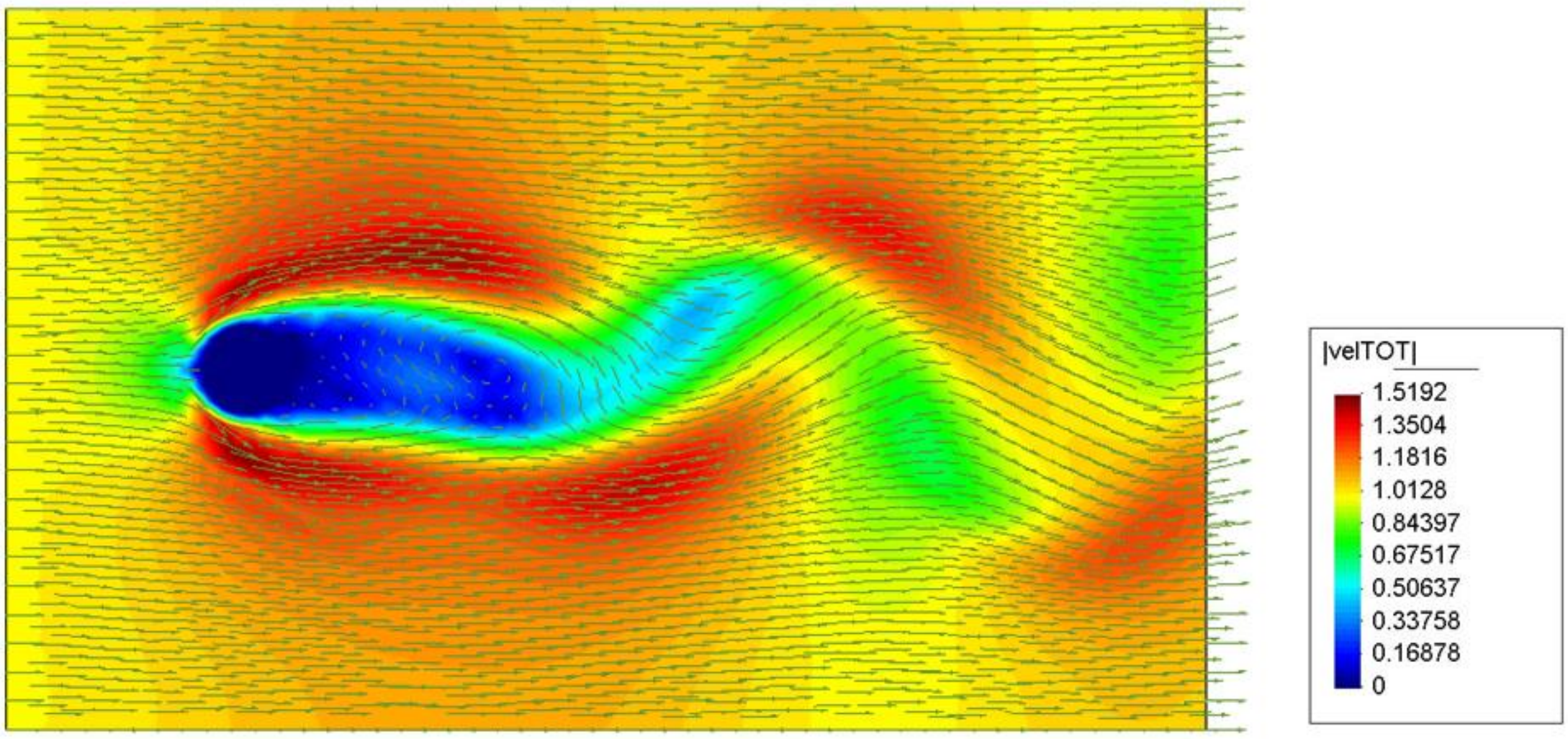

Figure 11: Flow around a circular cylinder: velocity vectors in the whole domain for Re = 90 (t = 212 s)

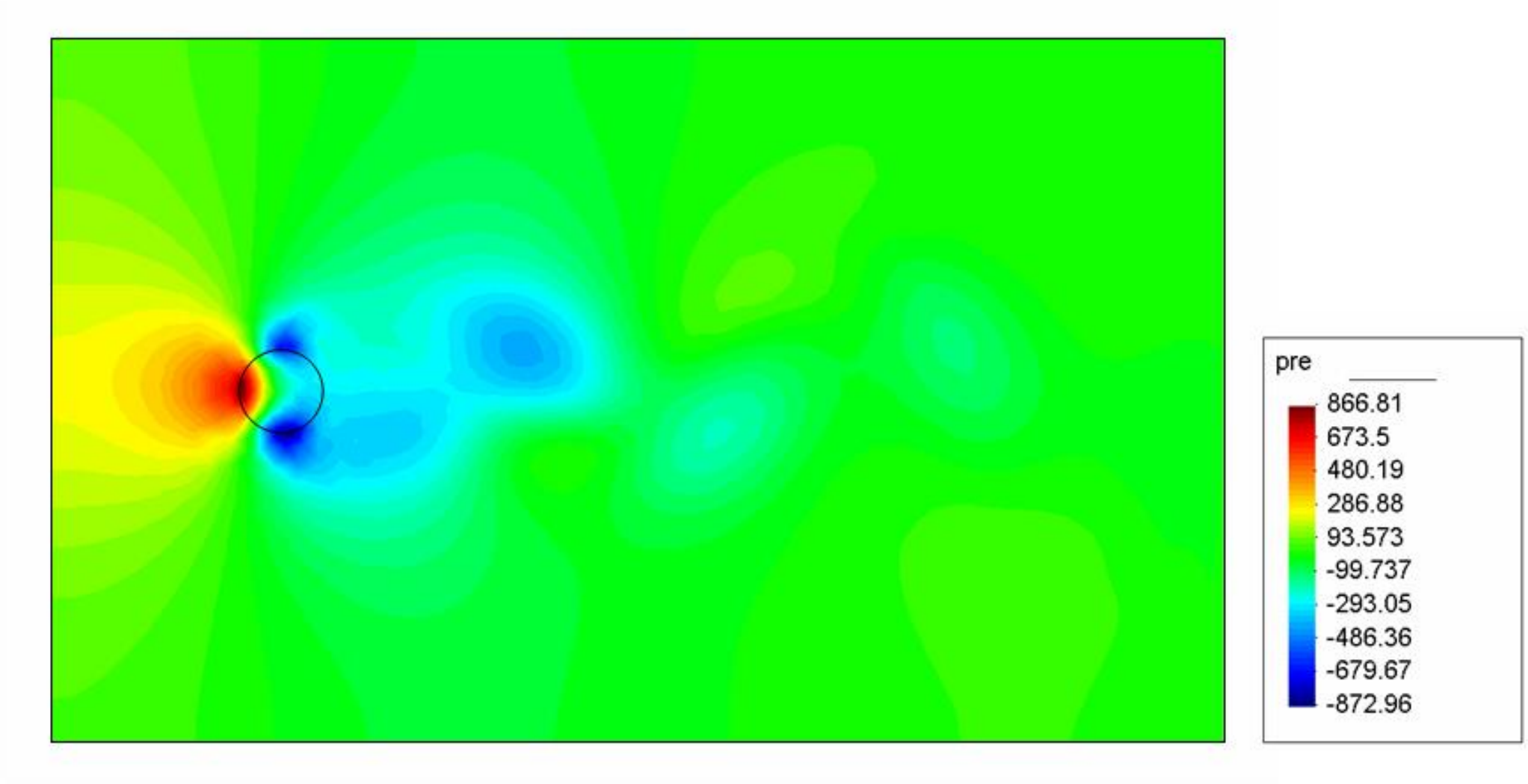

Figure 12: Flow around a circular cylinder: pressure values in the whole domain for Re = 90 (t = 212 s)

For Re = $10^3$: The periodic detachment of the eddies at the rear wake continues with a period of 5 s. Figures 13 and 14 depict the velocity field and pressure values in the whole domain for t = 212 s. Separation of the rear wake is observed at about θ = 80º (Figure 15) as observed experimentally [80,81].

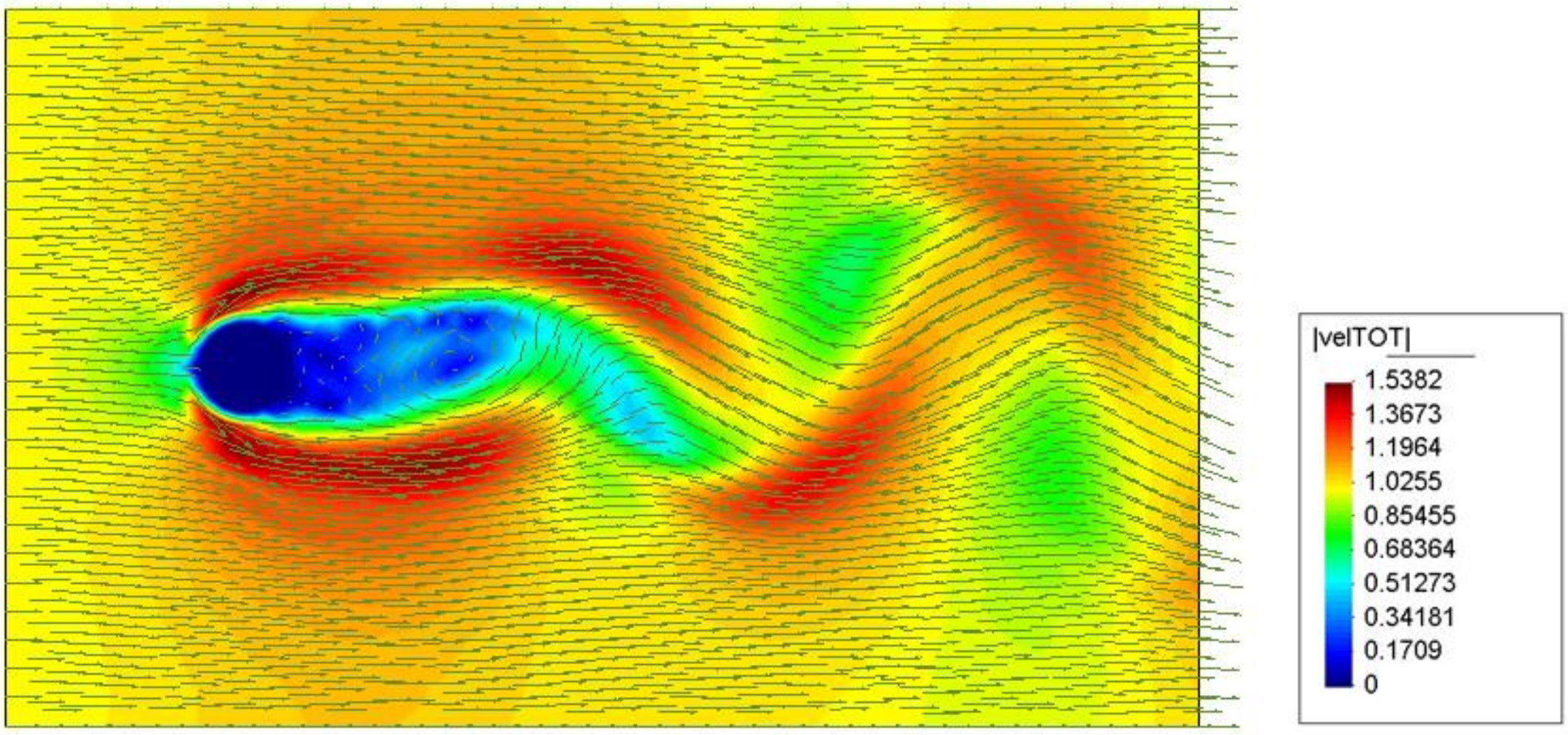

Figure 13: Flow around a circular cylinder: velocity vectors in the whole domain for Re = $10^3$ (t = 212 s)

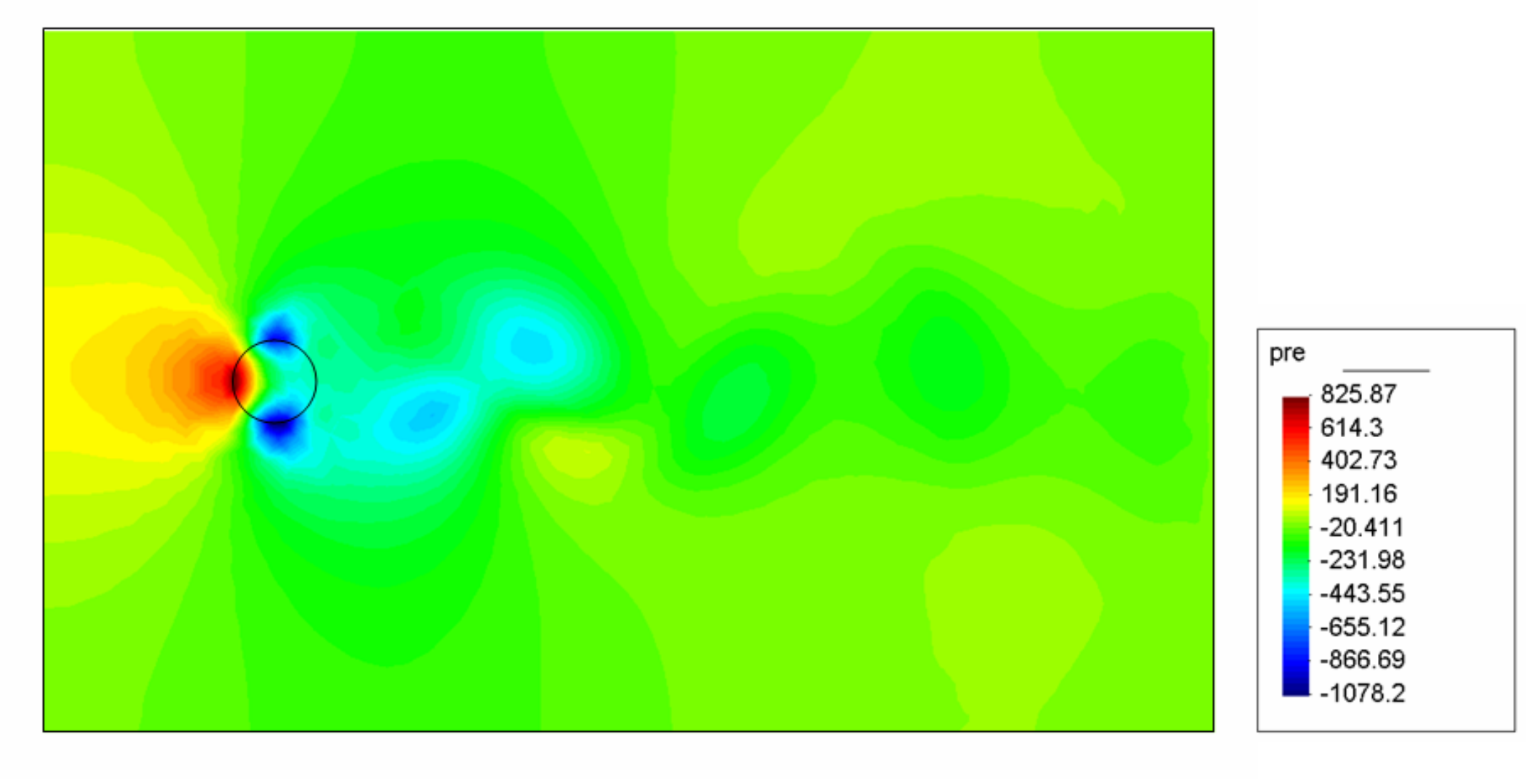

Figure 14: Flow around a circular cylinder: pressure values in the whole domain for Re = $10^3$ (t = 212 s)

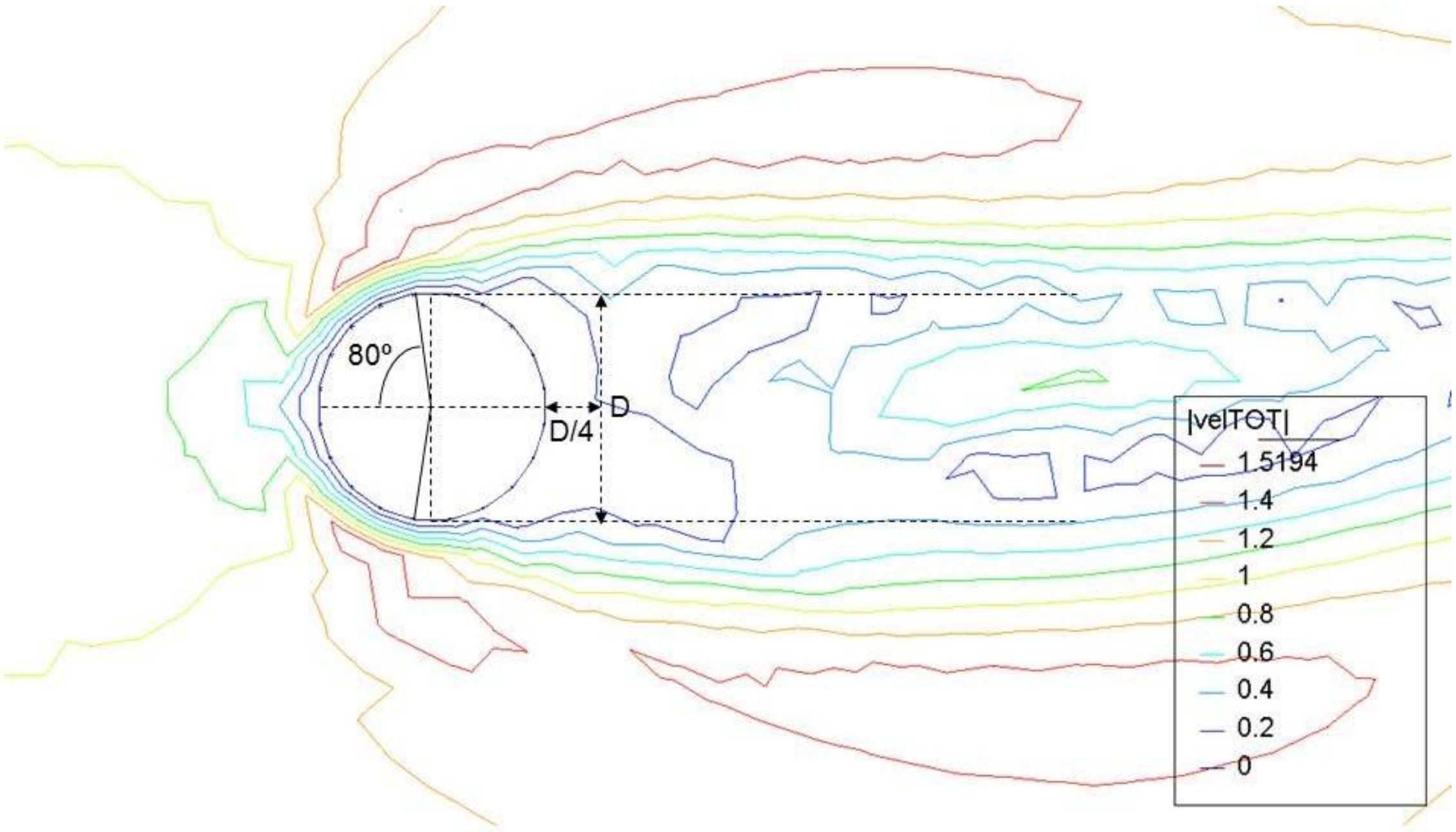

Figure 15: Flow around a circular cylinder: detailed velocity contours around the cylinder for Re = $10^3$ (t = 212 s)

For Re = 4 $10^5$: The periodic detachment of the eddies at the rear wake remains unchanged with a period of 5 s. Figures 16 and 17 depict the velocity vectors and pressure values in the whole domain for t = 212 s. Separation of the rear wake is observed at about θ = 120º (Figure 18) as observed experimentally [80,81]. Comparison of Figures 15 and 18 clearly show how the wake width becomes narrower for Re = 4 $10^5$ [80,81].

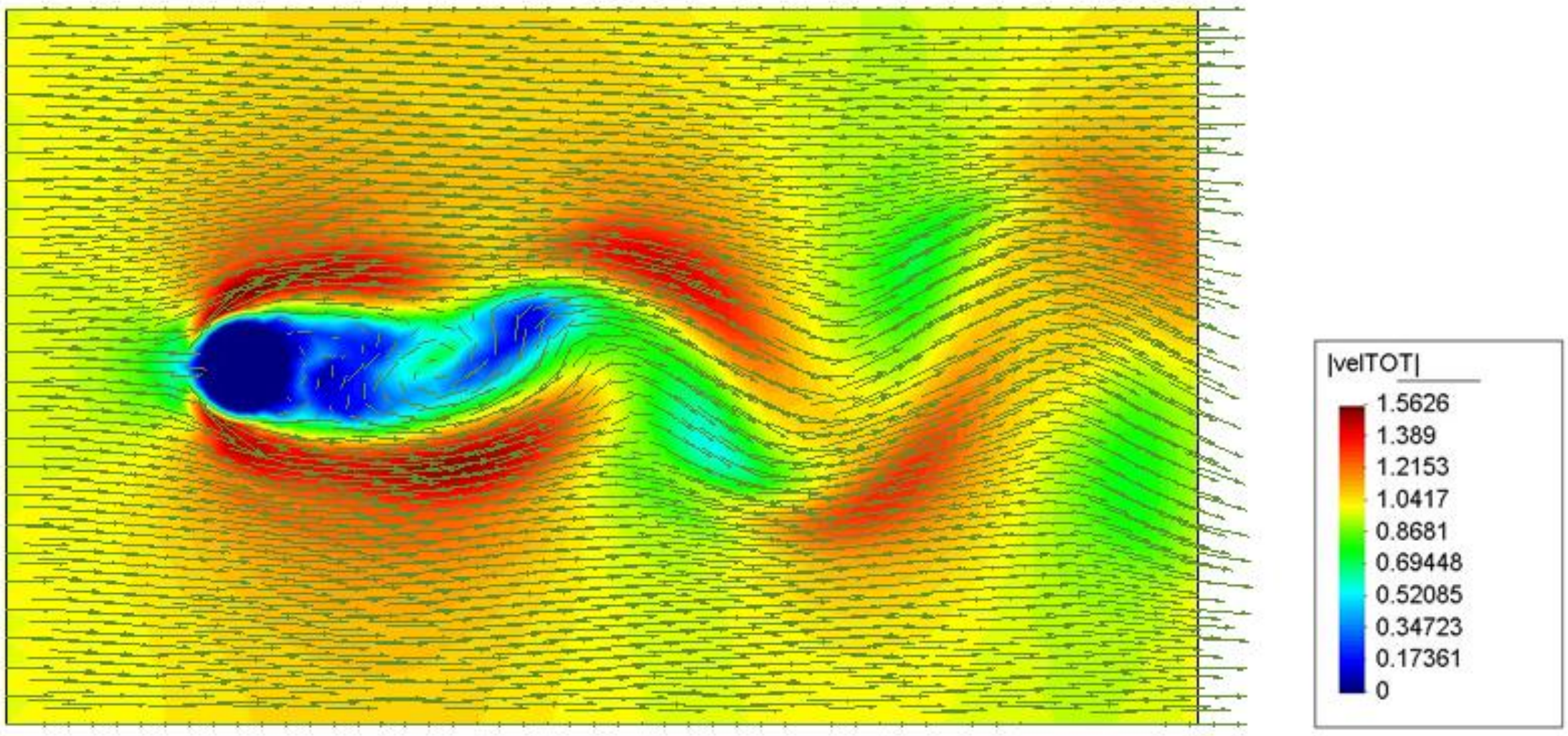

Figure 16: Flow around a circular cylinder: velocity vectors in the whole domain for Re = 4 $10^5$ (t = 212 s)

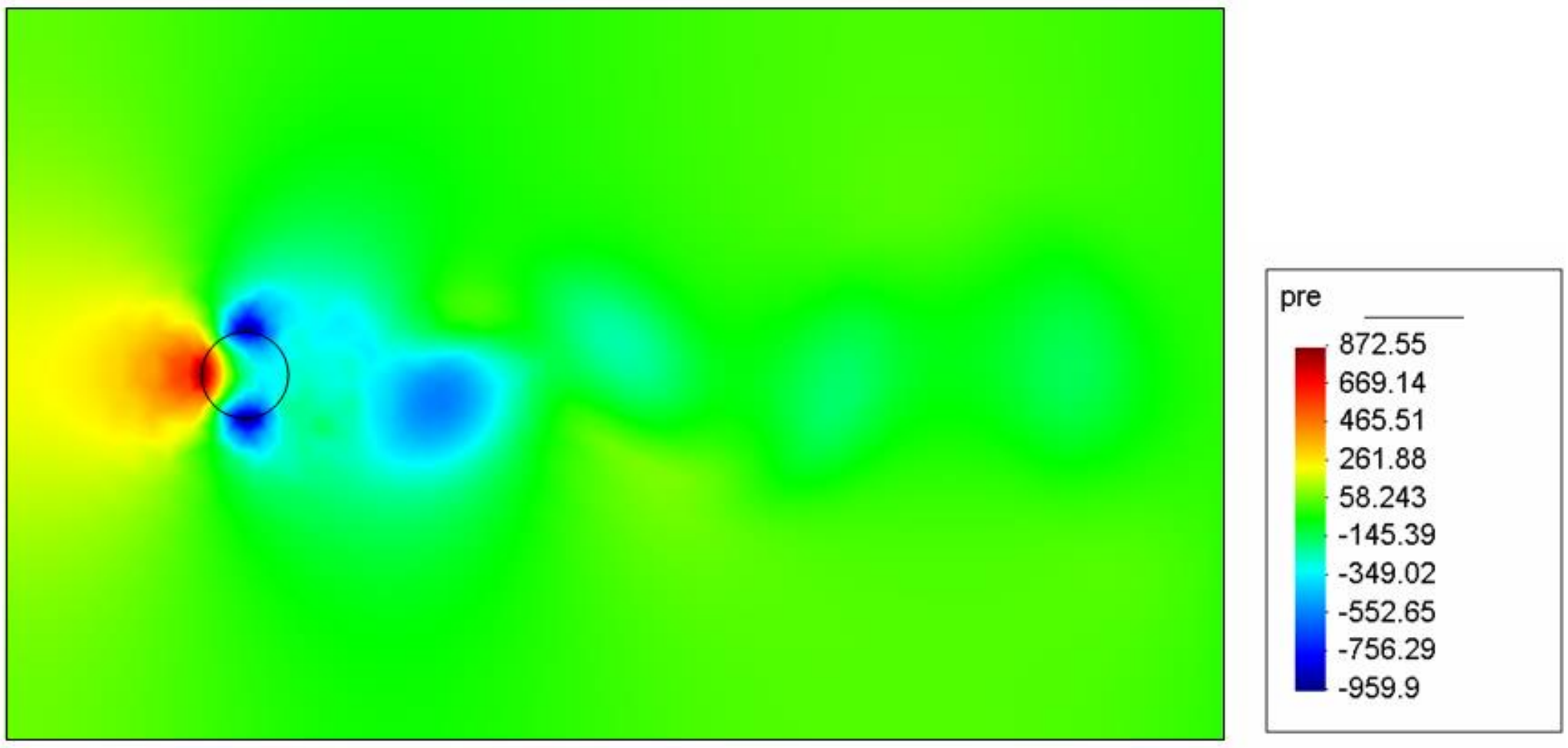

Figure 17: Flow around a circular cylinder: pressure values in the whole domain for Re = 4 $10^5$ (t = 212 s)

A characteristic feature of the vortex street generated at the rear of the cylinder is its geometric similarity. This feature of the vortex street suggests that there should be analytical relations between the Reynolds number (related to the lateral dimension of the obstacle, the flow velocity and the fluid viscosity) and parameters characterizing the flow, such as the Strouhal number, the drag coefficient and other parameters describing the vortex street [85]. Roshko [72-73] experimentally investigated laminar and turbulent wakes behind cylinders of different cross sections and observed geometric similarity among all vortex streets. From his data, he derived empirical features of vortex streets that are Reynolds-number independent, and he postulated a *universal* Strouhal number related to the wake width. In addition, the vortex-shedding process is slightly affected by other phenomena (e.g. oblique shedding modes, transition modes, Kelvin–Helmholtz instability, boundary-layer effects at the obstacle) although the main characteristics of the vortex street are not strongly affected by them [85]. Thus, the complexity of the process has hindered the development of an analytical description of the problem and consequently, numerous models have been proposed during the last decades.

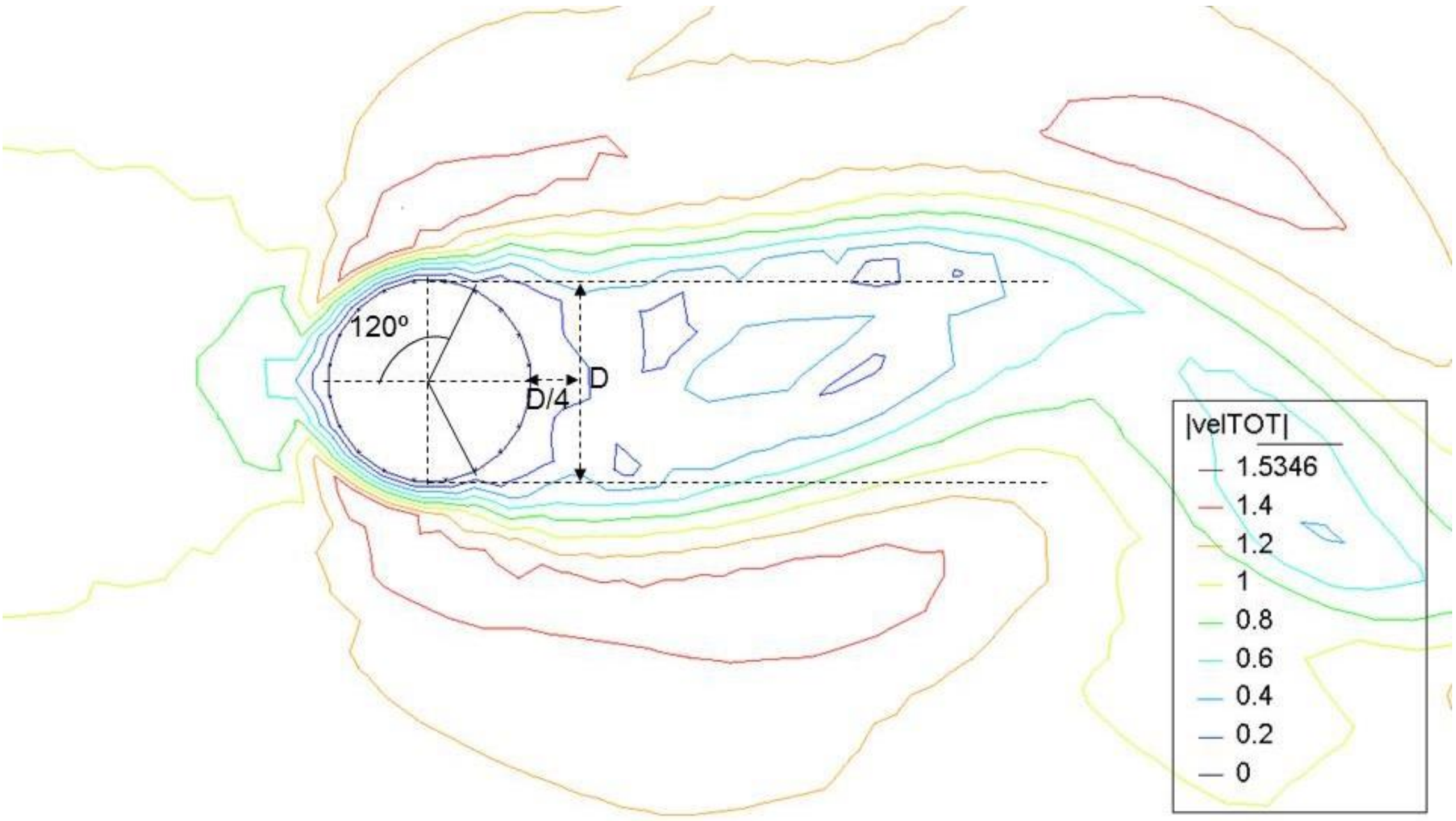

Figure 18: Flow around a circular cylinder: detailed velocity contours around the cylinder for Re = 4 $10^5$ (t = 212 s)

In this numerical example the Strouhal number dependency on the Reynolds number, $St = \frac{D}{T\,v_\infty} = f(Re)$, has been analysed and compared with some empirical models [73,82,83]. The results depicted in Figure 19 show that the proposed numerical approach agrees well with the results found in the literature, especially for Reynolds numbers larger than 160 [73,82,83]. For Re < 160 the proposed model seems to overestimate the empirical models proposed by Roshko [73], Ponta and Aref [82] and V. Strouhal [83], however the agreement is reasonably good comparing with the experimental results given in [84].

These observed differences in the results are mainly due to the fact that to accurately capture the vortex structures either very fine meshes must be considered or a special treatment for turbulence must be incorporated in the numerical model [86-87]. Nevertheless, the numerical treatment of turbulence at different scales falls out of the scope of the present study. In addition to that, experimental analysis and theoretical modelling of the Reynolds number dependency as well as the development of coherent structures, are still subject to research nowadays.

Finally, a numerical analysis of the pressure distribution on the cylinder and subsequent computation of the drag coefficient for an inviscid flow are carried out.

Figure 20 shows the computed values of the pressure coefficient around the cylinder, $c_p$, for different values of the Reynolds number. These numerical values are compared with the theoretical pressure coefficient predicted by the inviscid theory (46). It is observed that as the Reynolds number increases the minimum of the numerical curves gets lower values. In the limit, for an inviscid fluid, the numerical result is almost coincident with the theoretical curve until detachment of the rear wake occurs (120° < θ < 130°), which is not predicted by the inviscid theory.

Once the pressure coefficient is known, the drag coefficient in the inviscid case may be calculated as [80,88]:

$$c_D = \int_0^\pi c_p \cos\theta \; d\theta \approx 1.17 \qquad (47)$$

Figure 21 shows the drag coefficient evolution in time for the inviscid case. While the theoretical solution yields a constant value of the drag coefficient, the numerical solution is oscillatory, as observed experimentally for Re > 30 [78,79]. For the stabilized numerical solution (after 150 s), the mean value of the oscillating solution differs from the theoretical value with a relative error of 0.0010 (0.1%).

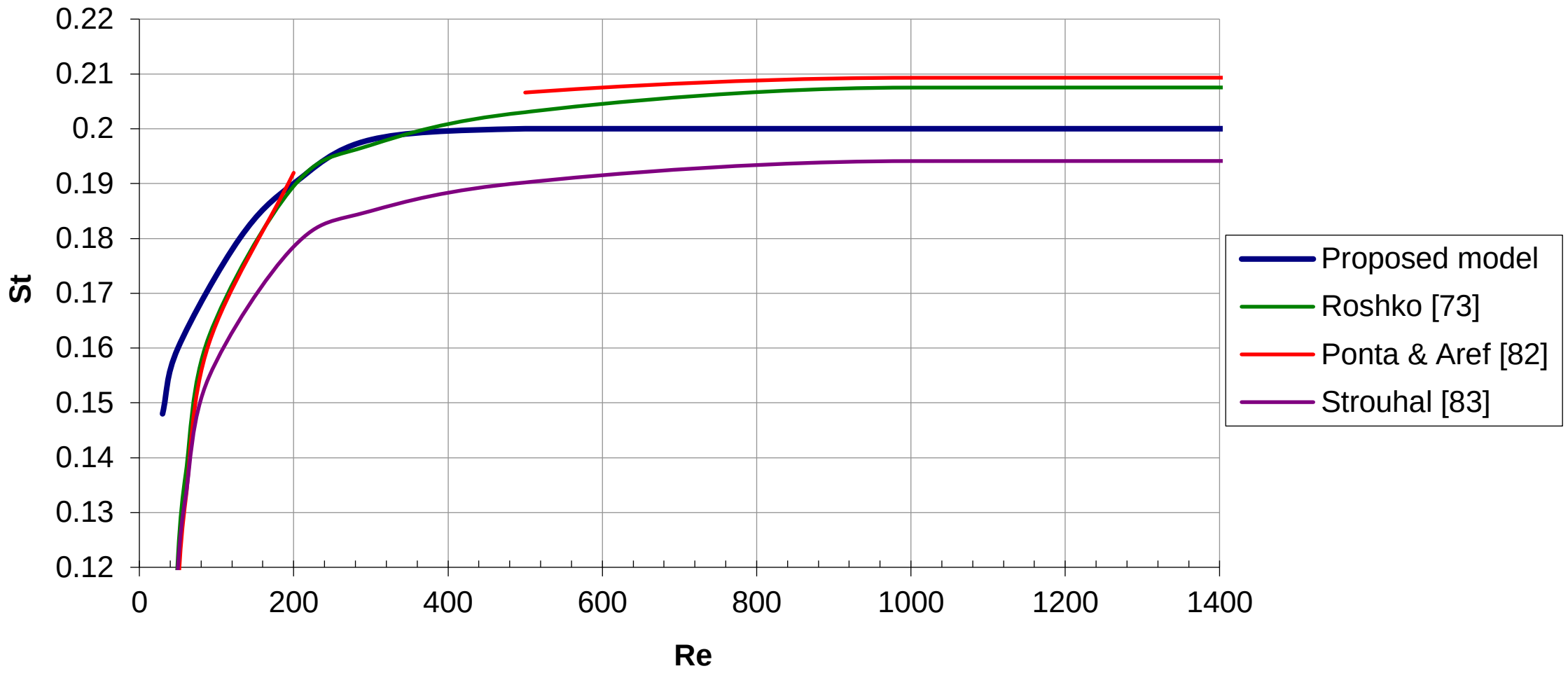

Figure 19: Flow around a circular cylinder: Strouhal number dependency on the Reynolds number

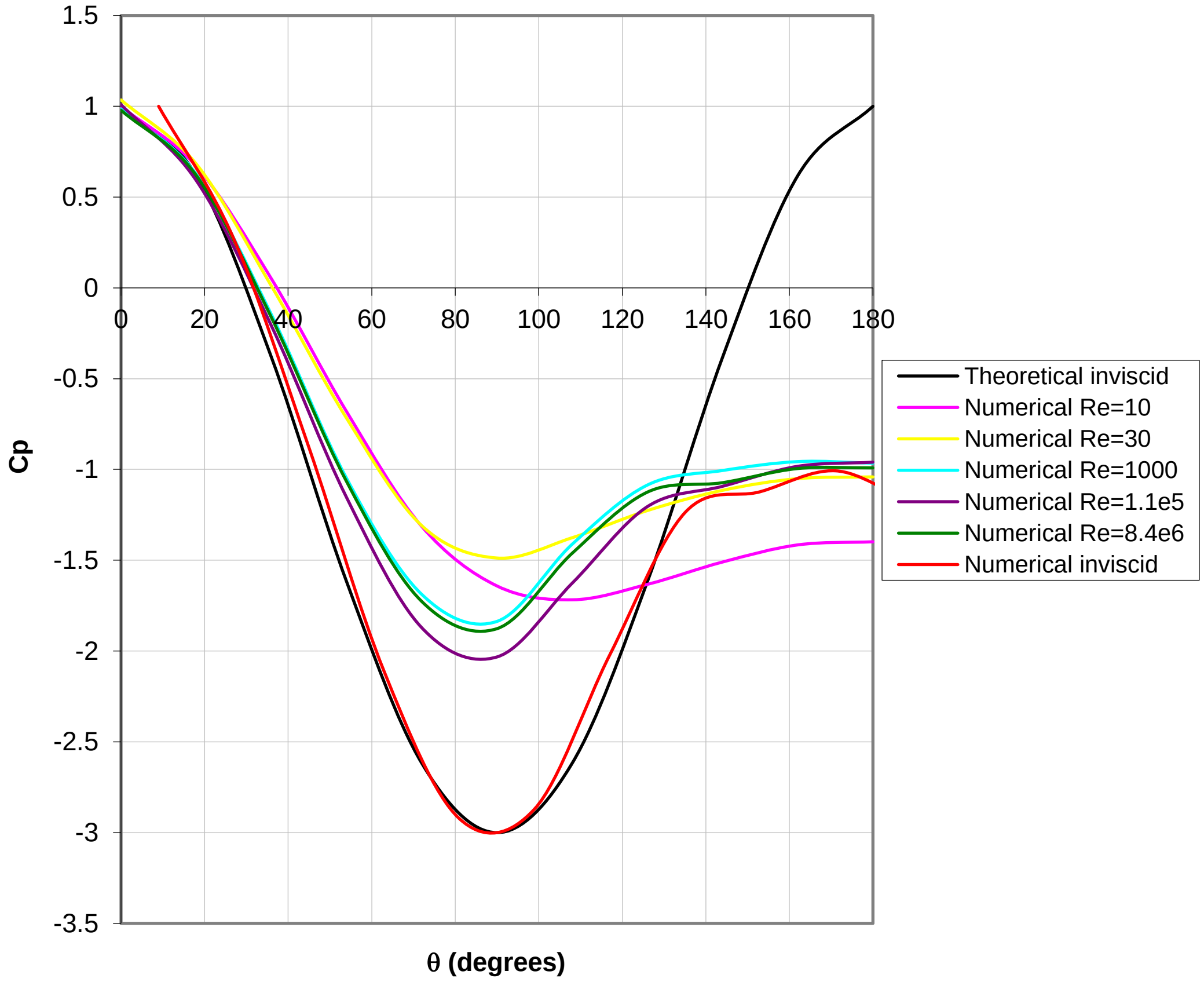

Figure 20: Flow around a circular cylinder: pressure coefficient on the cylinder for different values of the Reynolds number

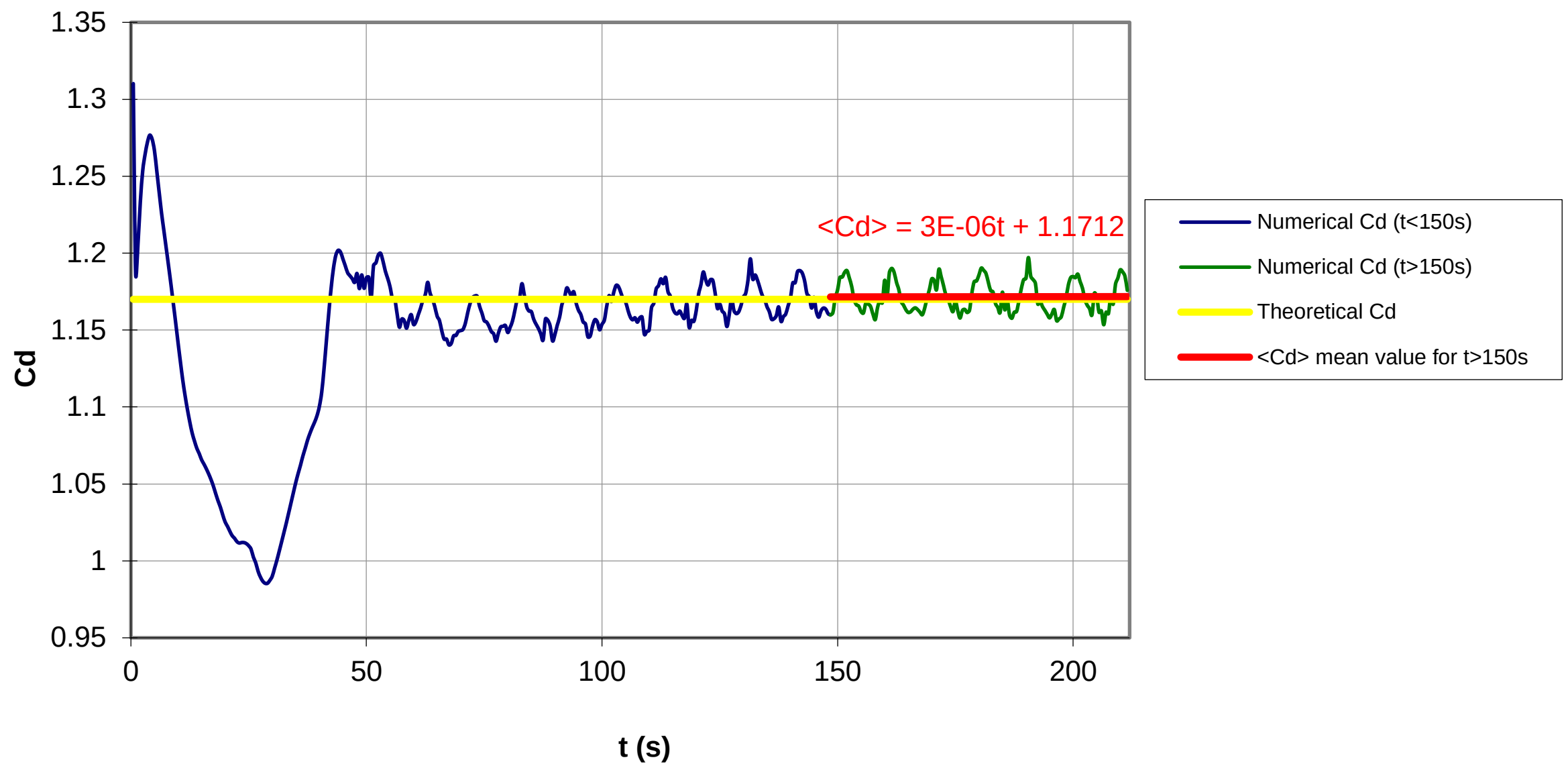

Figure 21: Flow around a circular cylinder: drag coefficient for an inviscid flow

### 4.2 Motion of a rigid square cylinder immersed in a rotational fluid flow

One of the main challenges for the numerical modelling of a rigid body moving inside a viscous fluid is maintaining the volume and shape of the solid throughout its movement. As explained in Sections 2-3, the solution adopted in this work is the combination of the level set technique and the two-step Taylor-Galerkin algorithm for tracking the fluid-solid interface. The characteristics exhibited by the two-step Taylor-Galerkin, minimizing oscillations and numerical diffusion, make this method suitable to accurately advect the solid domain avoiding distortions at its boundaries. Illustrating this important feature of the proposed model is the aim of the numerical example presented herein.

This problem consists of a solid square cylinder of D = 0.2 m immersed in a rotating viscous fluid whose initial velocity is $v_n = 0$ (normal component) and $v_t = 5\pi r$ m/s (tangential component), being r the distance to the center of the domain (see Figures 22 and 23). The boundary conditions are: $v_n = 0$ along the four sides of the domain and p = 0 at the middle point of each side (Figure 22).

A non-structured mesh of 11322 linear triangles (5728 nodes) is used for the computation (Figure 24). The parameters used in the computation are taken as follows: length and height of the fluid domain are L = 1 m and H = 1 m respectively (Figure 22); the fluid is assumed to be of newtonian type with viscosity $\mu = 1$ Pa s and density $\rho_f = 1000$ kg/m$^3$. The center of the solid square cylinder is placed at the center point of the domain (0.5 m from the left side of the domain and 0.5 m high, as shown in Figure 22) and its density is $\rho_s = 1000$ kg/m$^3$. Neither the solid position nor its velocity is restricted, thus both are left free to evolve throughout the simulation. The time-step used for the calculation is $\Delta t = 2.83\ 10^{-4}$ s.

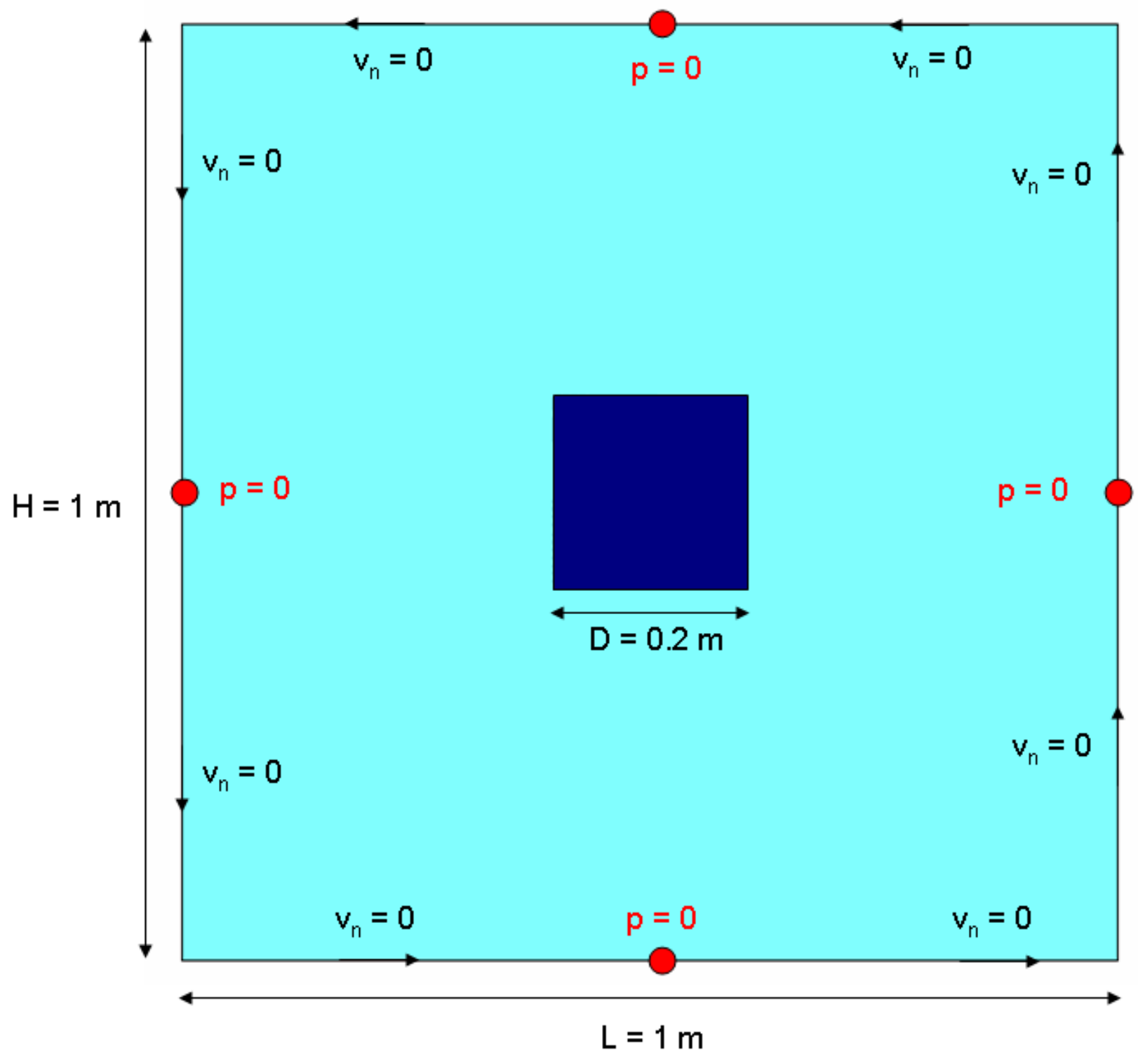

Figure 22: Rigid square cylinder immersed in a rotational fluid flow: problem layout and boundary conditions.

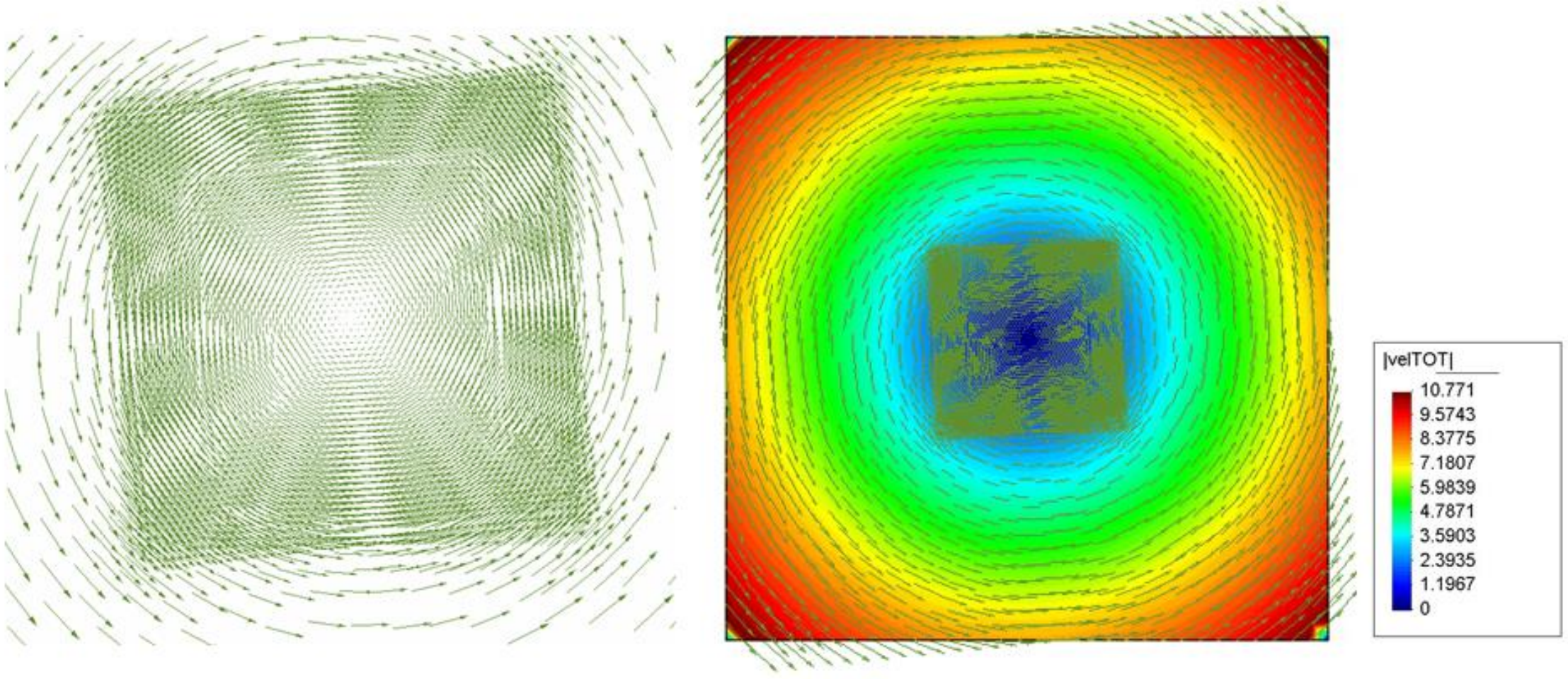

Figure 23: Rigid square cylinder immersed in a rotational fluid flow: initial conditions for velocity in the whole domain (right) and detailed velocity vectors in the solid domain (left).

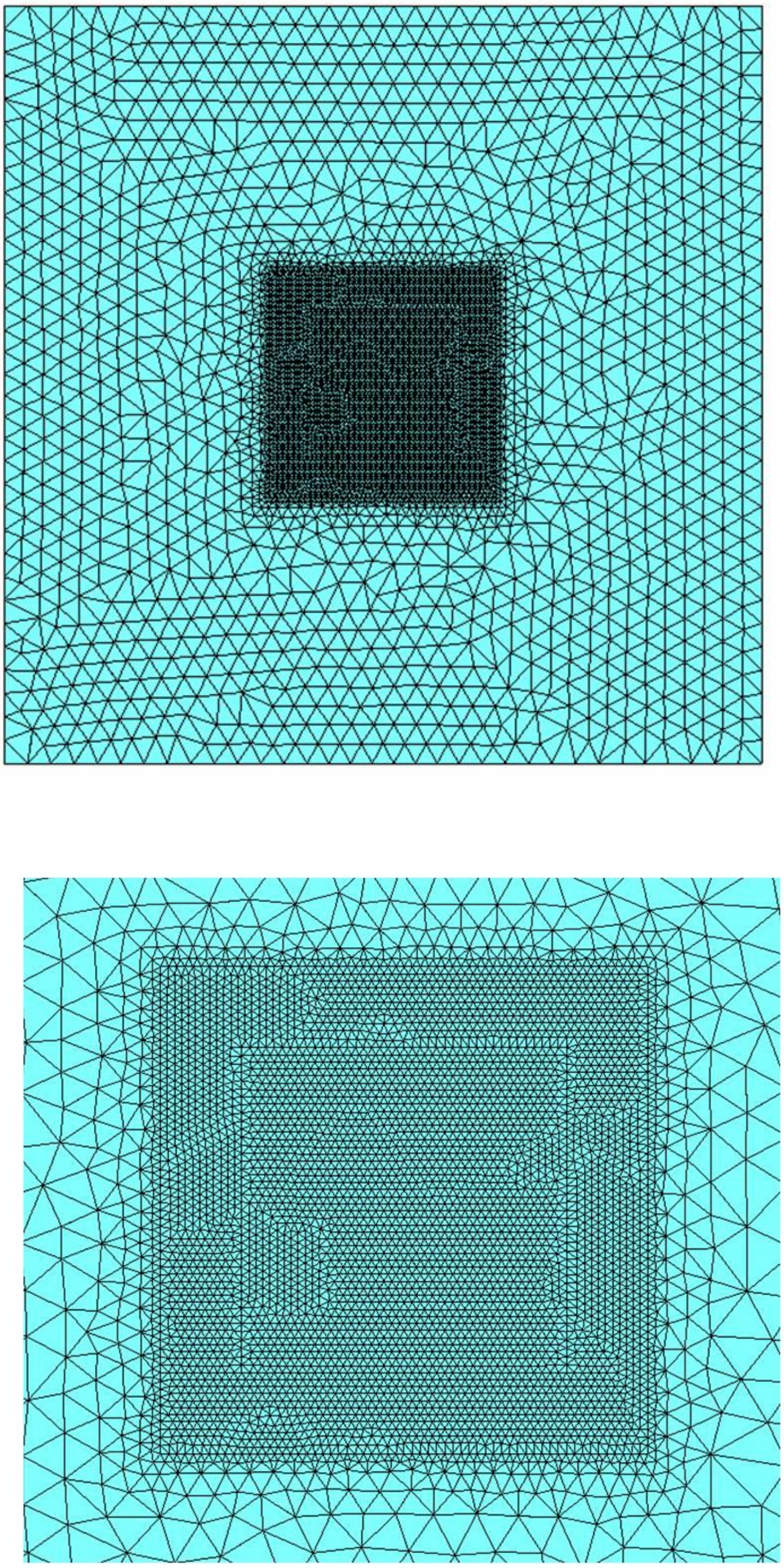

Figure 24: Rigid square cylinder immersed in a rotational fluid flow: computational mesh in the whole domain (top); detailed computational mesh inside the solid domain (bottom).

Figure 25 shows the position of the fluid-solid interface for 8 different times during a total revolution of the solid about its center of mass. It can be observed that the shape and volume of the body are well preserved after a very high number of iterations ($5\ 10^3$ iterations). Even though the rigid body is free to move inside the domain, the absence of instabilities allows the solid to spin about its center without either translating or distorting.

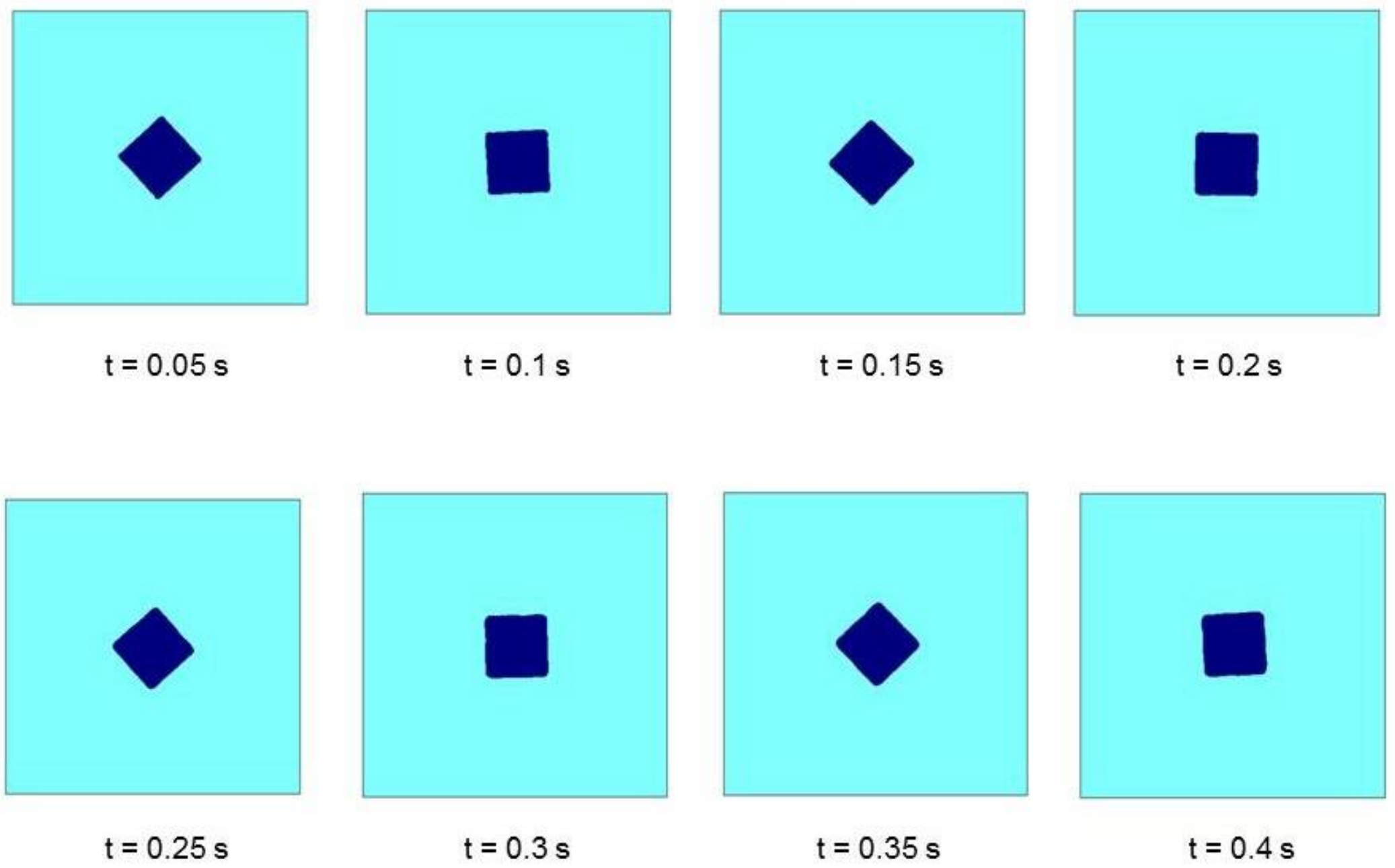

Figure 25: Rigid square cylinder immersed in a rotational fluid flow: evolution of the fluid-solid interface during 0.4 s (5 $10^3$ iterations)

According to the initial velocity field, $v_t = 5\pi r$ m/s, the angular velocity is $\omega = 5\pi$ rad $s^{-1}$. Thus theoretically, in t = 0.4 s the solid must perform one complete rotation about its center of mass, which is in accordance with the obtained numerical result (see Figure 25). Velocity vectors and pressure distribution after one total rotation of the solid are shown in Figures 26, 27 and 28. The CPU time (2.5GHz-4Gb) for this simulation is 145 s.

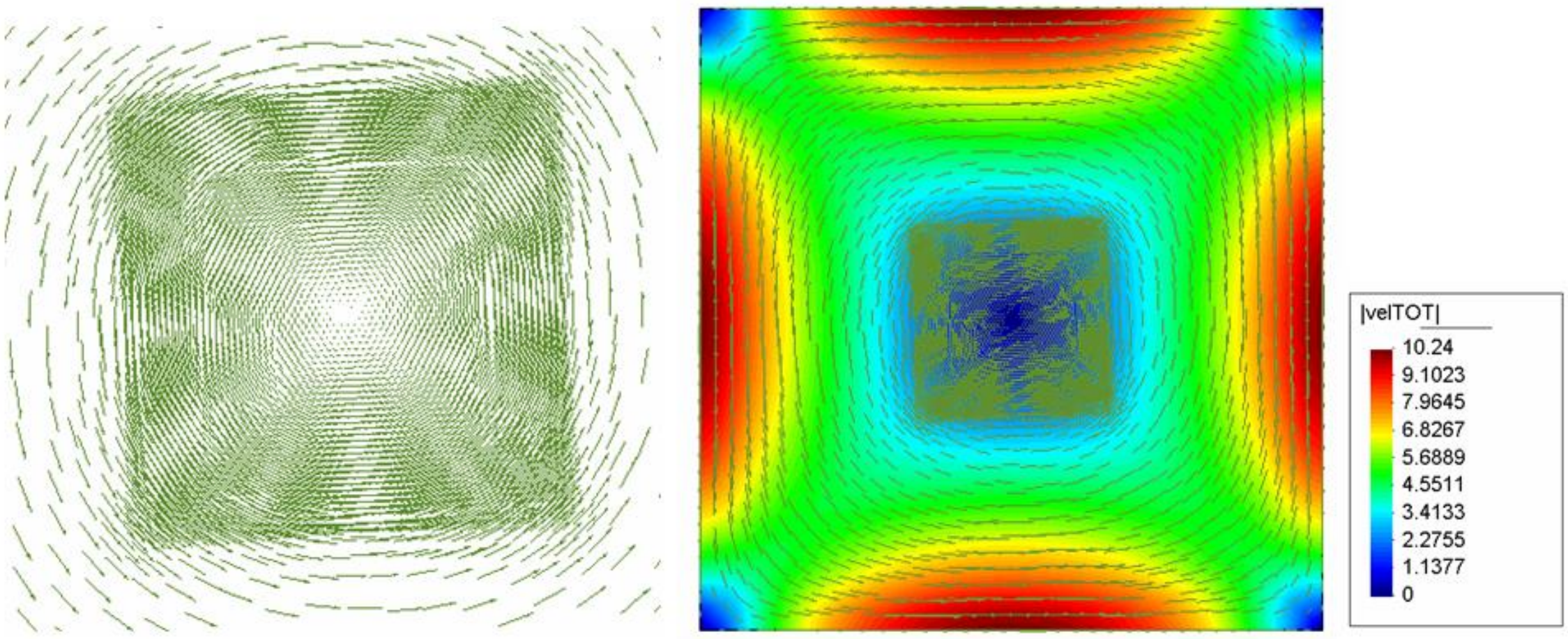

Figure 26: Rigid square cylinder immersed in a rotational fluid flow: velocity vectors for t = 0.25 s in the whole domain (right) and inside the solid domain (left).

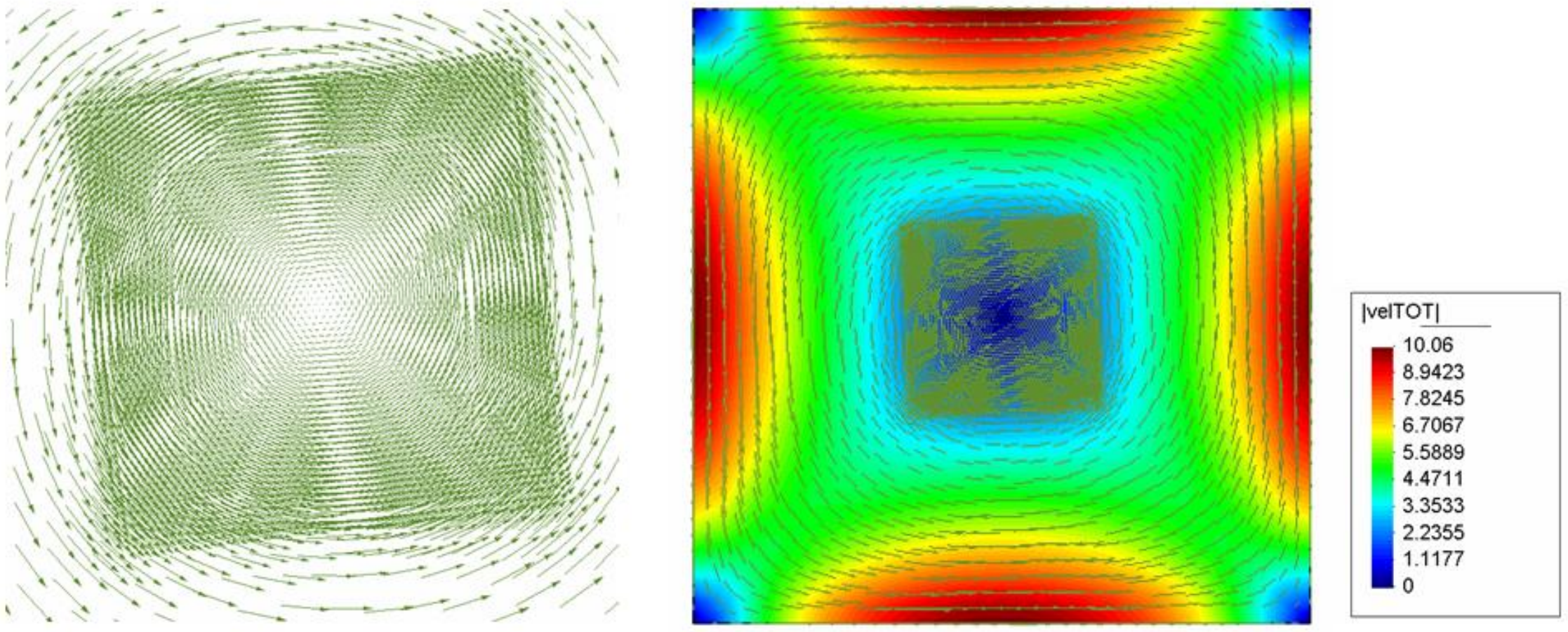

Figure 27: Rigid square cylinder immersed in a rotational fluid flow: velocity vectors for t = 0.4 s in the whole domain (right) and inside the solid domain (left).

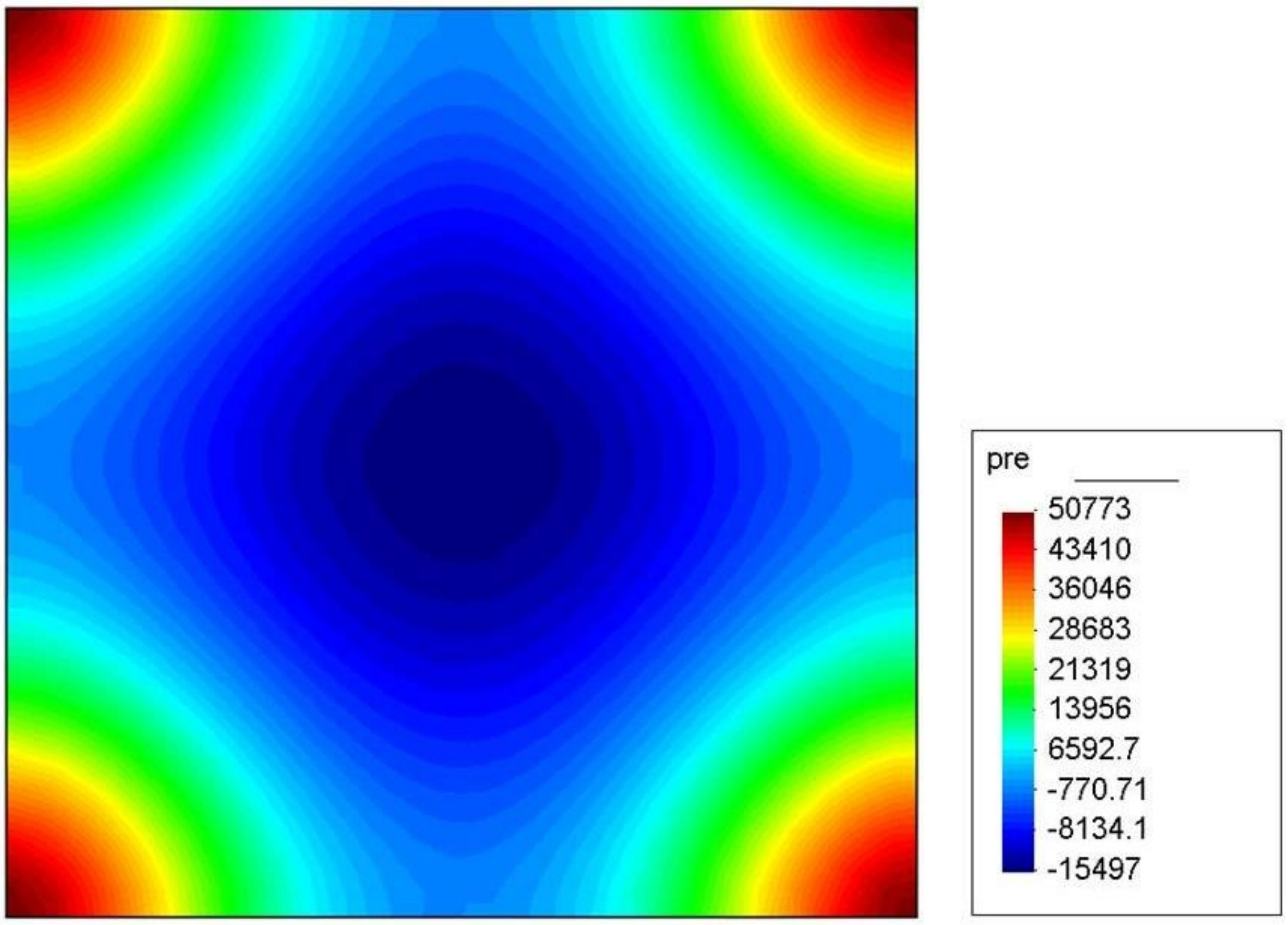

Figure 28: Rigid square cylinder immersed in a rotational fluid flow: pressure distribution for t = 0.4 s

### 4.3 Particle settling in a viscous fluid

Finally, to assess the dynamical behaviour of a submerged body inside a viscous flow, the gravity-driven motion of a cylinder in a viscous fluid is considered. The aim of this example is to validate the performance of the model under dynamic conditions, in a case in which experimental and numerical solutions are available.

*4.3.1. Settling of a circular cylinder with $\rho_s > \rho_f$*

The layout of the problem is sketched in Figure 29. A solid circular cylinder of diameter D = 0.05 m and density $\rho_s$ = 7800 kg/m$^3$ is immersed in a viscous fluid of newtonian type with viscosity $\mu$ = 8 Pa s and density $\rho_f$ = 1200 kg/m$^3$ and subject to gravity g = 9.8 m/s$^2$. The length and height of the fluid domain are given by L = 1.4 m and H = 2.43 m respectively. The center of the cylinder is placed at 0.7 m from the left side of the domain and 1.62 m high. No movement restrictions are imposed to the solid. No-slip boundary conditions are assumed along the four boundaries and pressure is prescribed as p = 0 in the top corners of the domain. Initial conditions for velocity are $v_x = v_y = 0$ in the whole domain, while hydrostatic initial conditions for pressure are considered.

A non-structured mesh of 9950 linear triangles (5073 nodes) is used for the computation (Figure 30). The size of mesh elements in the finest area (solid domain) is $h_{min}$ = 0.008 m while $h_{max}$ = 0.05 m in the coarsest part of it. The time-step used for the computation is $\Delta t = 1.08\ 10^{-4}$ s.

The proposed model allows the computation of the velocity field and pressure distribution in the domain induced by the rigid body motion. As shown in Figure 31 for t = 1.26 s no oscillations or instabilities are present even in the area surrounding the cylinder. Figure 32 shows the vertical velocity distribution along a diametric horizontal line across the domain once the terminal velocity is achieved.

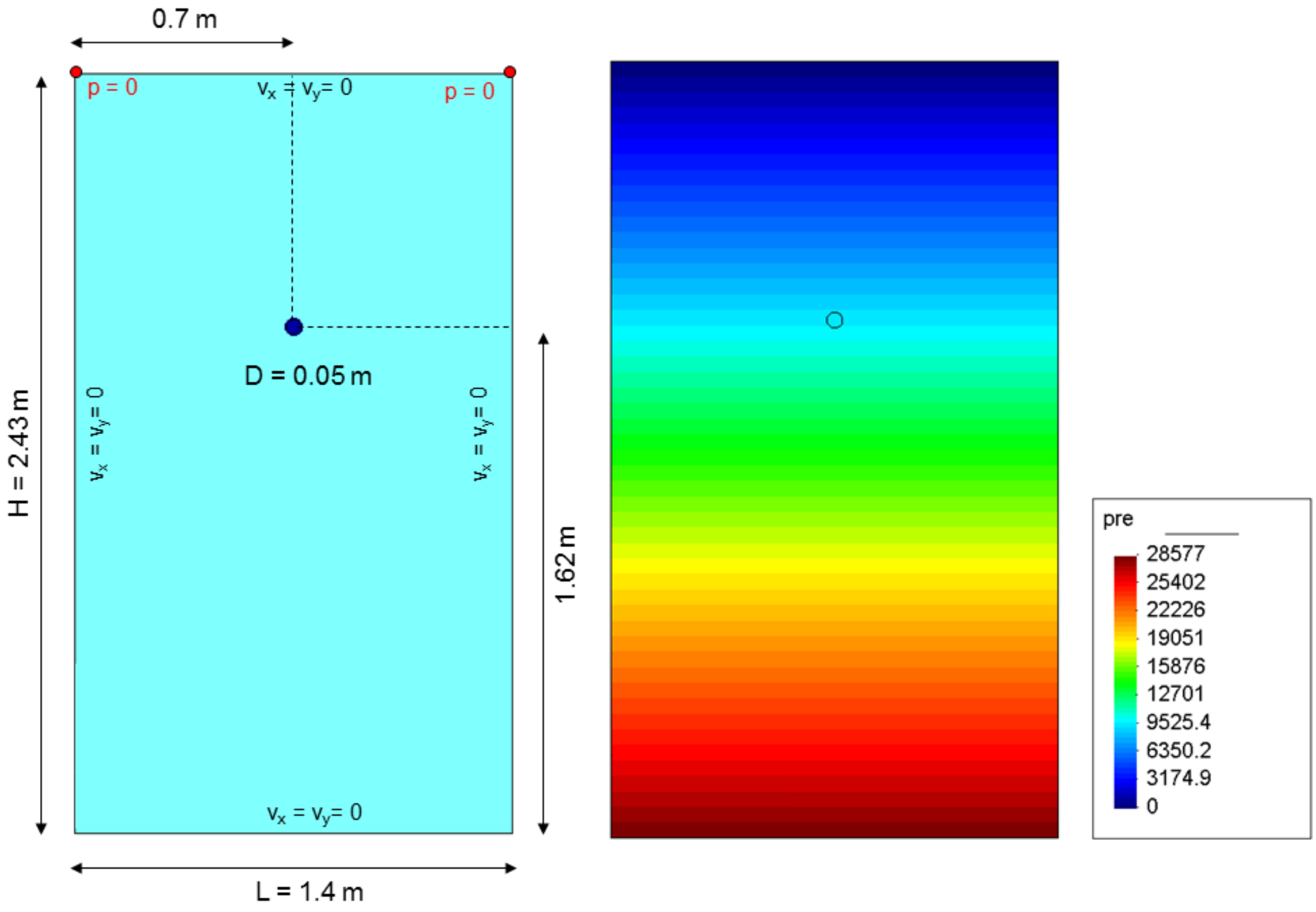

Figure 29: Settling of a circular cylinder: problem layout and initial conditions for velocity and pressure.

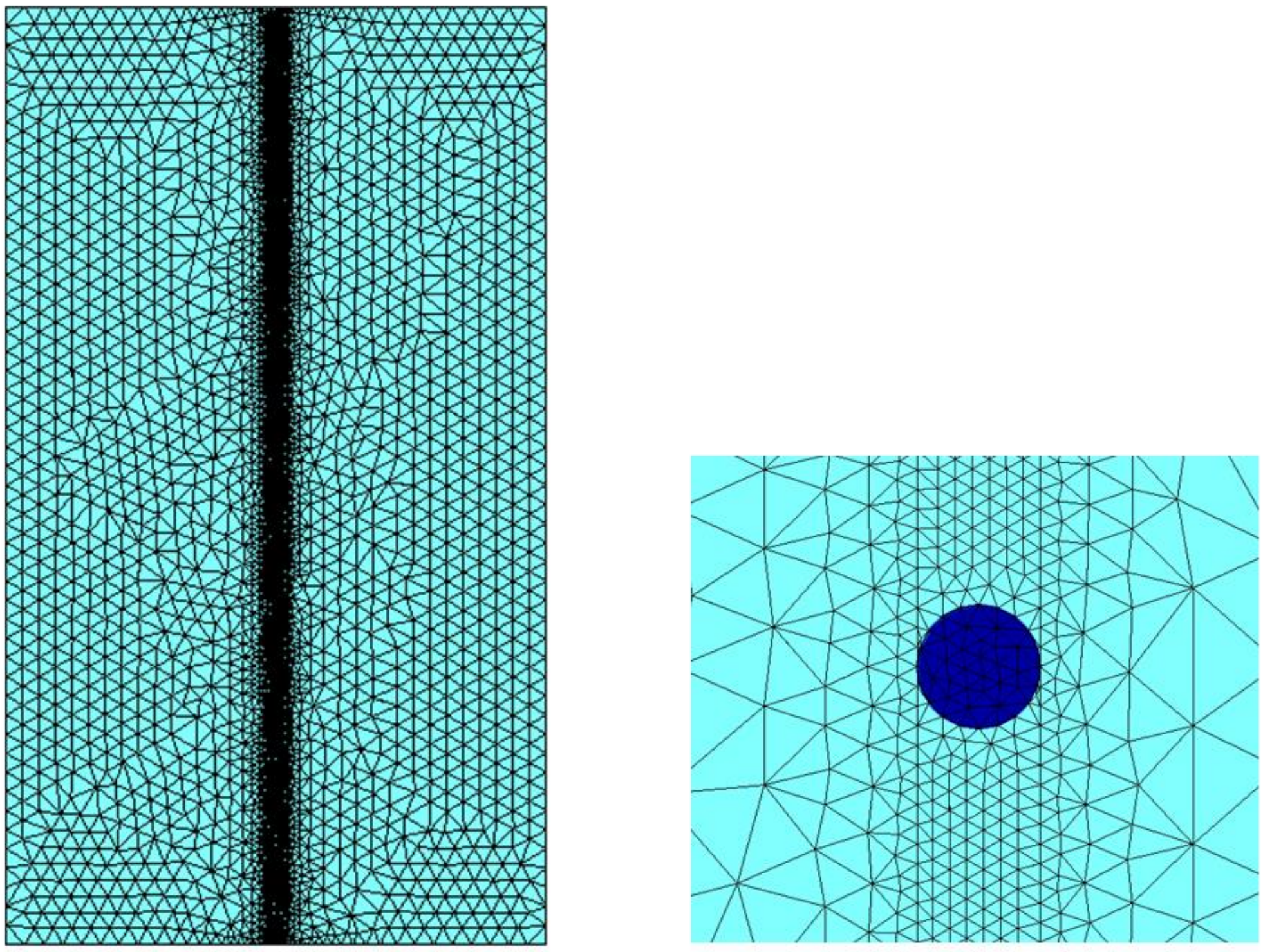

Figure 30: Settling of a circular cylinder: Computational mesh in the whole domain (left) and detailed mesh inside the solid domain (right).

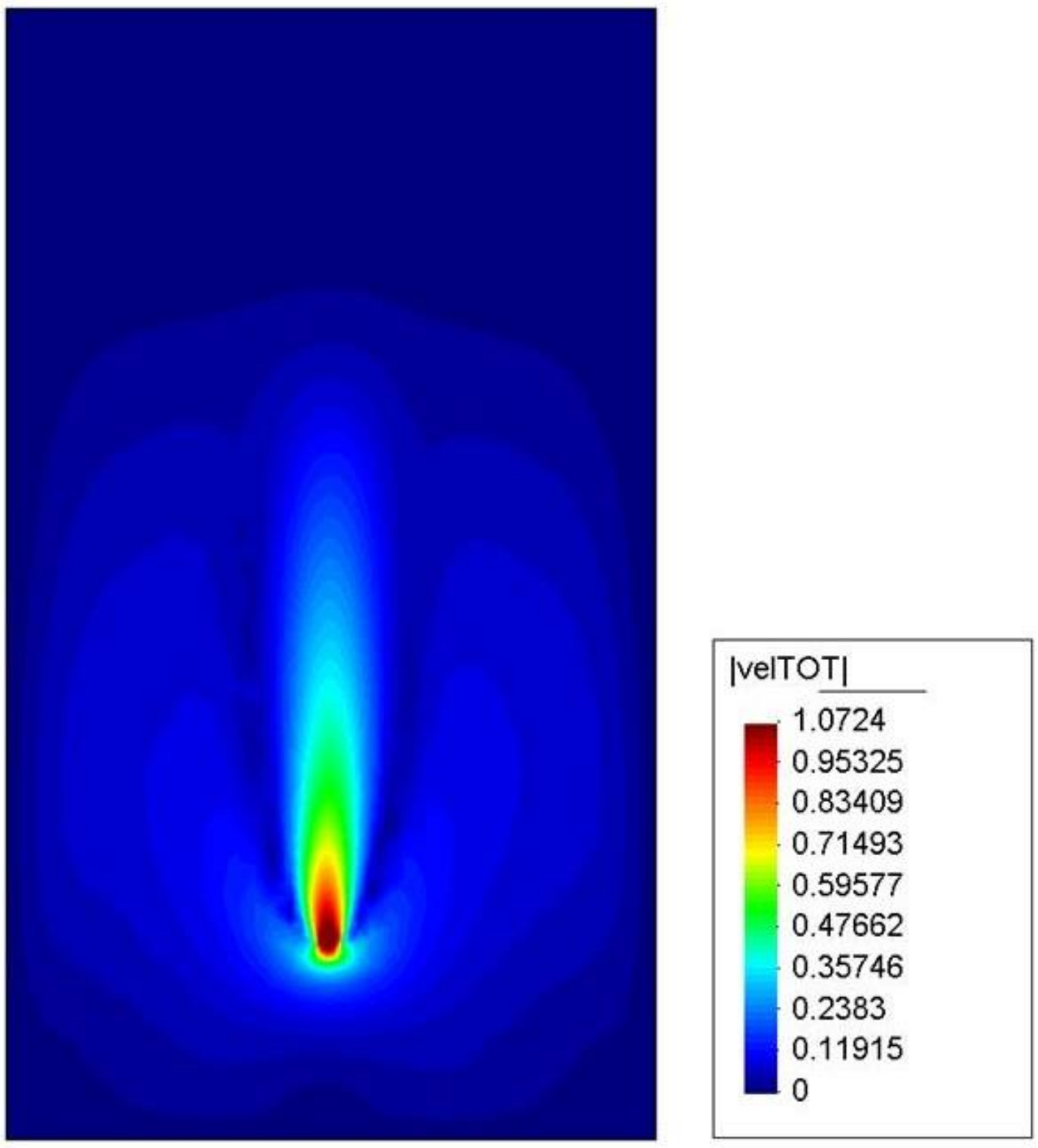

Figure 31: Settling of a circular cylinder with $\rho_s > \rho_f$: velocity distribution for t = 1.26 s

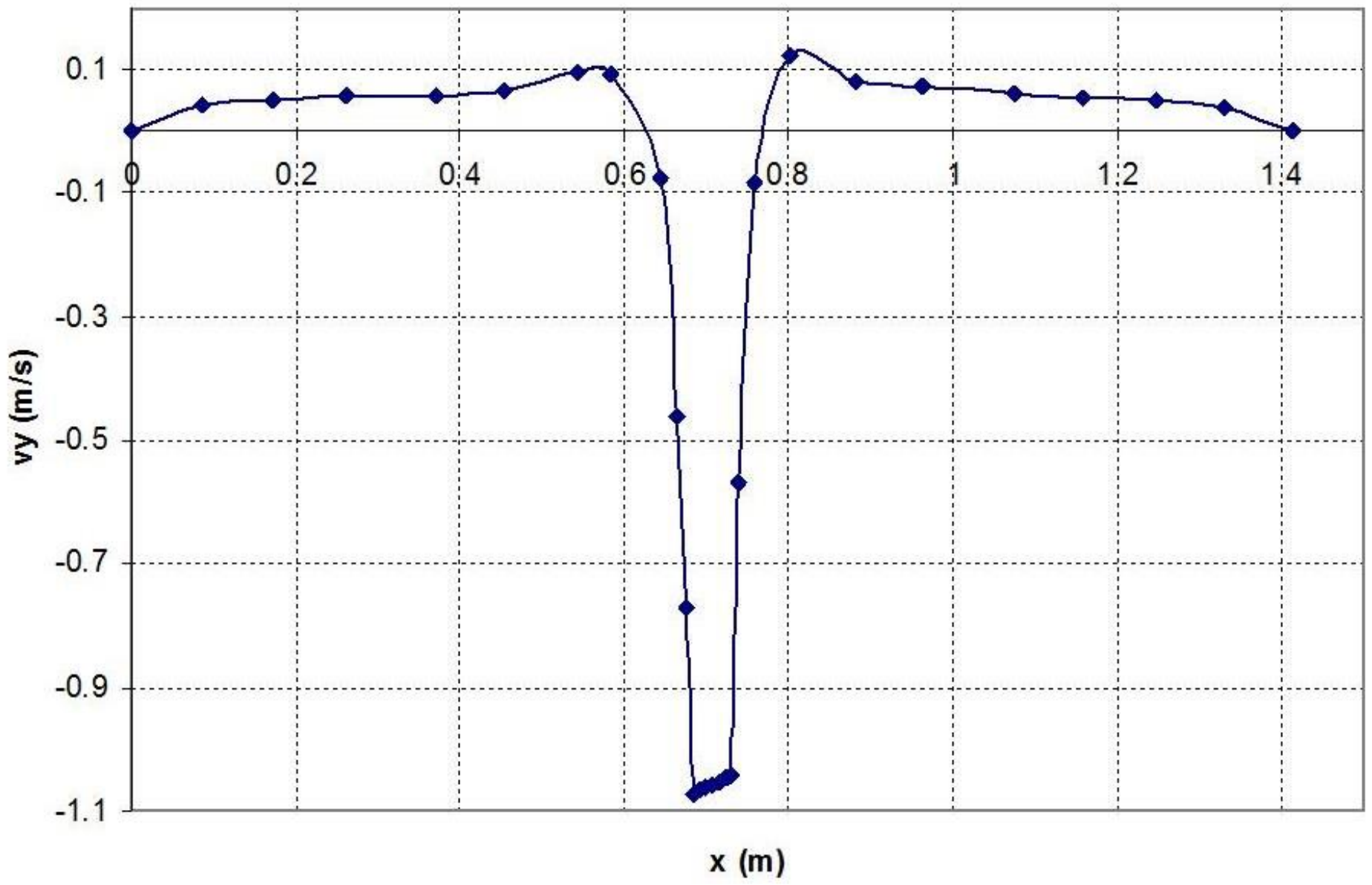

Figure 32: Settling of a circular cylinder with $\rho_s > \rho_f$: vertical velocity distribution along a diametric horizontal line across the domain for t = 1.26 s

Since no movement restrictions are imposed to the cylinder, the rigid body is free to move in all directions. Evolution of the solid center of mass is plotted in Figure 33. The rigid body follows a vertical path whose deviation from the vertical line is of the order of the minimum mesh element size, $h_e$ (relative error of $\varepsilon_r = 0.0249$). The terminal settling velocity can be derived from Figure 34. For the time interval between t = 0.3 s and t = 1.3 s the vertical velocity curve can be approximated by a linear function with a correlation factor $R^2$ = 1. The slope of the linear function is the value of the terminal velocity, $v_{terminal}$ = 1.079 m/s

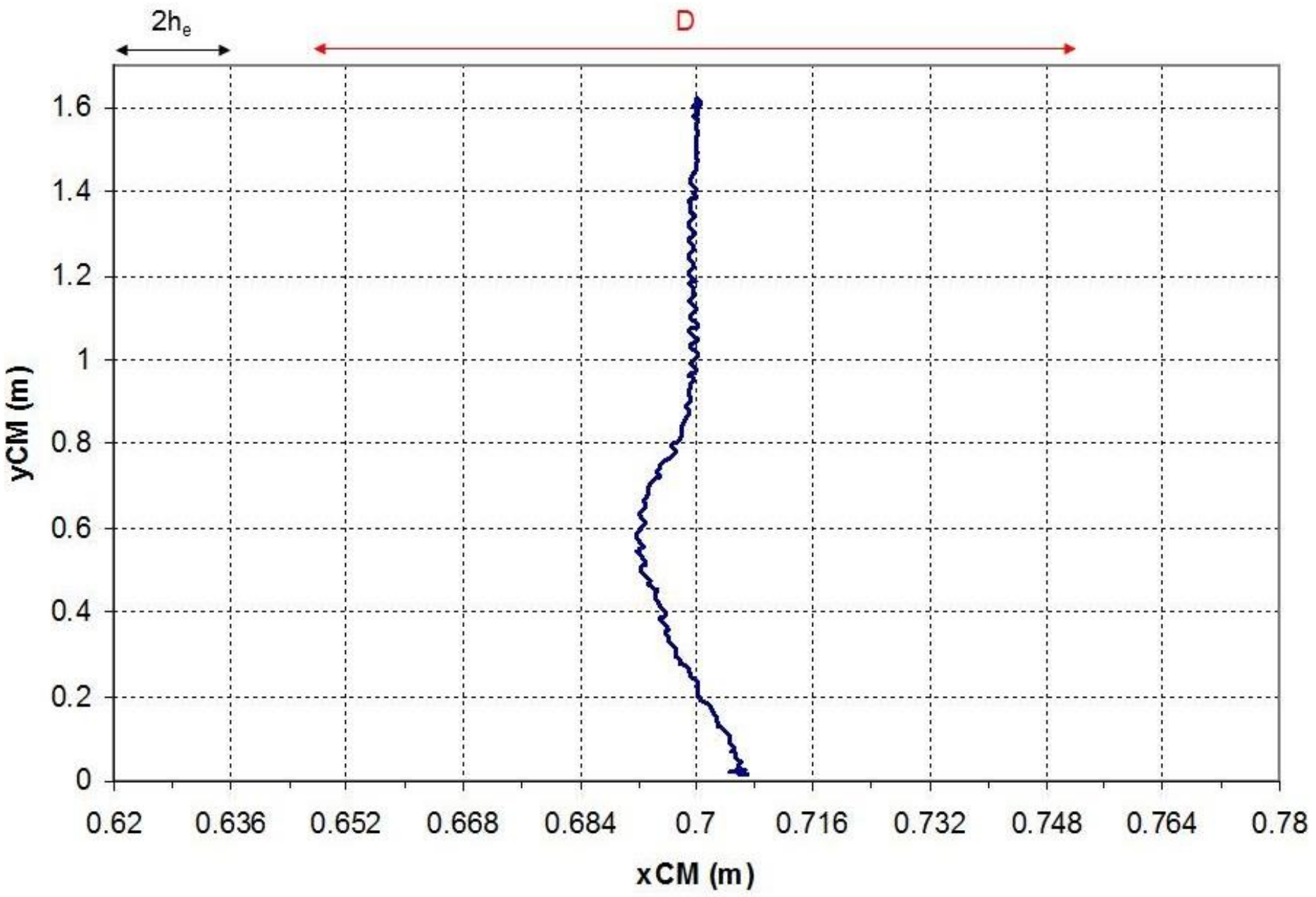

Figure 33: Settling of a circular cylinder with $\rho_s > \rho_f$: path followed by the solid center of mass

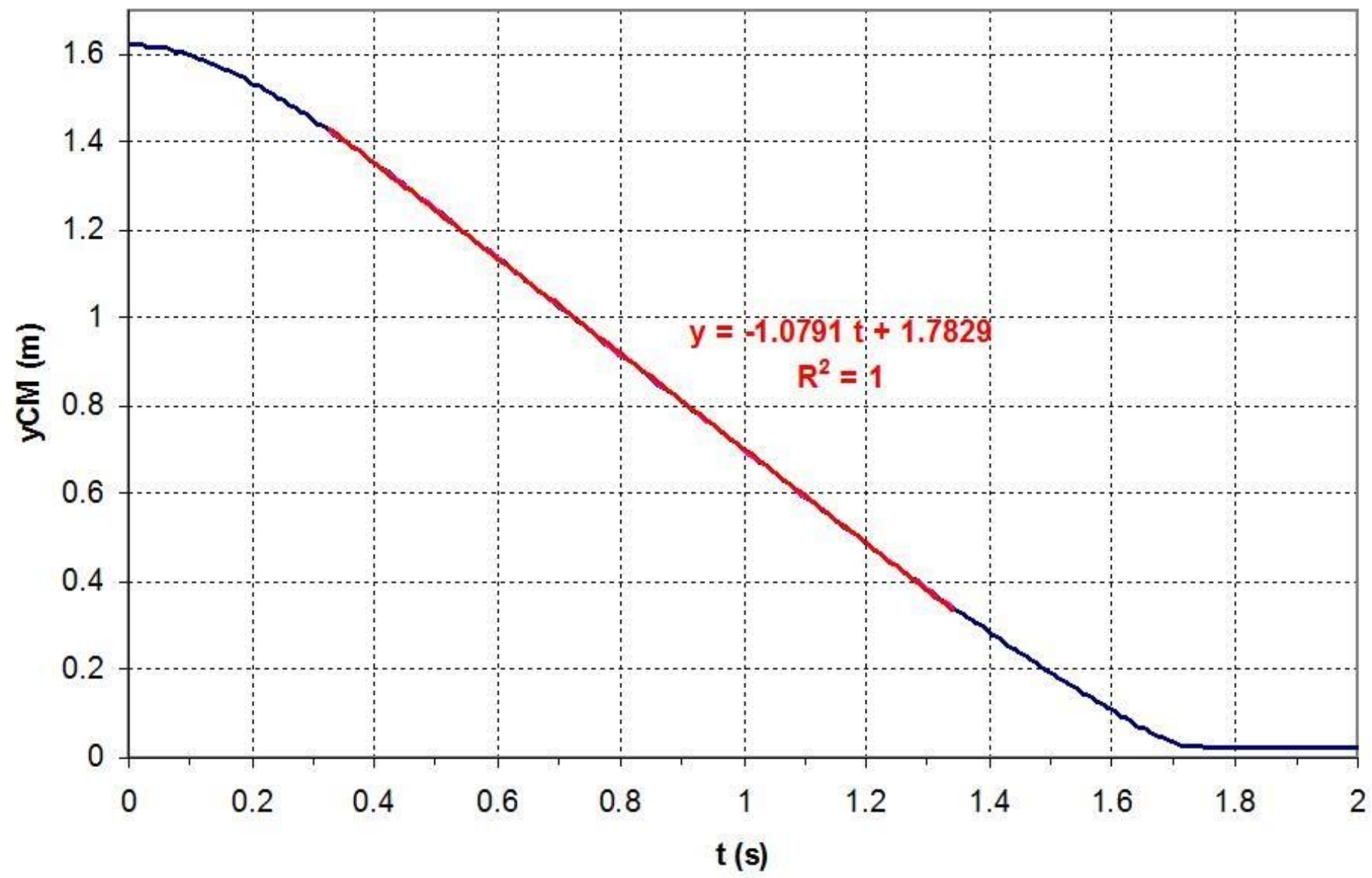

Figure 34: Settling of a circular cylinder with $\rho_s > \rho_f$: evolution in time of the y-coordinate of the solid center of mass and linear regression.

Some experimental correlations have been proposed by several authors in order to establish the relationship between $C_D$ and Re [81,89]. For 4 < Re < 10, the following expression can be considered:

$$C_D = 6.8052\ Re^{-0.3755} \tag{48}$$

For the computed settling velocity, $v_{terminal}$ = 1.079 m/s the Reynolds number is Re = 8.09 and therefore, according to (48) the resulting value of the drag coefficient is $C_D$ = 3.103.

This problem has been previously analysed experimentally and numerically by [89] and the experimental-based values for the terminal settling velocity and the drag coefficient are $v_{terminal}$ = 1.067 m/s and $C_D$ = 3.12 respectively, which confirms the good performance of the numerical model proposed herein.

In order to quantify the deformation of the solid and, therefore, the error of the model, the strain-rate tensor, $\mathbf{D}(\mathbf{u})$, in the L2-norm is computed for $\mathbf{x} \in S(t)$ [90,91]

$$||\mathbf{D}(\mathbf{u})||^2_{L2} = \sum_{S(t)}(D_{11}^2 + 2D_{12}^2 + D_{22}^2)h_e^2 \tag{49}$$

being $D_{ij} = \frac{1}{2}\left(\frac{\partial u_i}{\partial x_j} + \frac{\partial u_j}{\partial x_i}\right)$ the components of the symmetric tensor $\mathbf{D}(\mathbf{u})$, and $h_e$ the element size.

To test the sensitivity of the proposed method with respect to the penalty parameter, the dependency of the rate of deformation tensor, $\mathbf{D}(\mathbf{u})$, in the L2-norm for $\mathbf{x} \in S(t)$ on the penalty parameter, $\lambda$, is analysed. This test will allow quantification of the model error as a function of the penalty parameter, $\lambda$.

Table 5 summarizes the results for this case study. As previously shown in [64-66], there exists an intimate coupling between $\lambda$ and $\Delta t$: the error saturates around $\lambda \approx 1/\Delta t$. This behaviour can be understood by the interpretation of the penalty term as a strong damping term on the velocity, introducing a characteristic time scale, which is of order $1/\lambda$. Attending the interpretation of $1/\lambda$ as a physical permeability an apparent good option would be choosing a very high value for $\lambda$

>> 1/Δt. Unfortunately, the time step restriction for the proposed explicit scheme prevents from doing so, and therefore, the only possible way for using $\lambda >> 1/\Delta t$ would be treating the penalization term with an implicit time discretization scheme [16,66]. The results presented in Table 5 illustrate this fact and justify the use of $1/\lambda \approx \Delta t$ in the present work, including this particular example for which $\lambda = 9.2\ 10^3$ and $\Delta t = 1.08\ 10^{-4}$ s. In Table 5 are also displayed the computational times for all tested cases, ranging from 313 s to 1830 s. For the case study presented in this section, the rate of deformation tensor, **D**(**u**), in the L2-norm is $6.88\ 10^{-3}$ and the CPU time (2.5GHz-4Gb) is about 387.48 s, showing the balance between accuracy and low computational cost of the proposed model.

| **λ** | **Δt** | **$\|\|D(u)\|\|_{L2}$** | **Simulation time (s)** | **CPU time (s)** |
|---|---|---|---|---|
| 1.0e5 | 1.97e-5 | 6.56e-3 | 3 | 1829.91 |
| 1.0e4 | 1.23e-4 | 7.32e-3 | 3 | 363.02 |
| 9.2e3 | 1.08e-4 | 6.88e-3 | 3 | 387.48 |
| 5.8e3 | 1.38e-4 | 5.57e-3 | 3 | 324.75 |
| 4.2e3 | 1.19e-4 | 2.09e-3 | 3 | 362.11 |
| 2.4e3 | 1.04e-4 | 2.21e-2 | 3 | 420.17 |
| 1.0e3 | 1.20e-4 | 2.79e-2 | 3 | 354.13 |
| 8.3e2 | 1.20e-4 | 6.66e-2 | 3 | 352.90 |
| 8.0e1 | 1.25e-4 | 0.49 | 3 | 325.98 |
| 8.0e0 | 3.80e-4 | 1.14 | 3 | 312.87 |

Table 5: Settling of a circular cylinder with $\rho_s > \rho_f$: sensitivity analysis with respect to the penalty parameter.

*4.3.2. Settling of a circular cylinder with $\rho_s < \rho_f$*

In order to test the performance of the model for a rigid body lighter than the fluid, a solid circular cylinder of diameter D = 0.05 m and density $\rho_s$ = 500 kg/m$^3$ subject to gravity g = 9.8 m/s$^2$ is considered. The problem geometry and initial and boundary conditions are the same as in the previous case (Figures 29-30). The fluid properties considered are: viscosity $\mu$ = 4 Pa s and density $\rho_f$ = 1200 kg/m$^3$. Same non-structured mesh of 9950 linear triangles (5073 nodes) is used for the computation (Figure 30). The time-step used in the calculation is $\Delta t = 1.69\ 10^{-4}$ s.

Velocity field in the domain induced by the rigid body motion is depicted in Figure 35 for t = 0.55 s. Again, no oscillations or instabilities are observed. Figure 36 shows the vertical velocity distribution along a diametric horizontal line across the domain once the terminal velocity is achieved.

The rigid body is free to move in all directions, since no movement restrictions are imposed. Evolution of the solid center of mass is plotted in Figure 37. The rigid body follows a vertical path whose deviation from the vertical line is of the order of the mesh size, $h_e$ (relative error of $\varepsilon_r$ = 0.0236). The terminal ascending velocity can be derived from Figure 38. For the time interval between t = 0.53 s and t = 1.70 s the vertical velocity curve can be approximated by a linear function with a correlation factor $R^2$ = 0.9999. The slope of the linear function is used to get the value of the terminal velocity, $v_{terminal}$ = 0.3936 m/s.

For t = 0.55 s, the calculated error given by (49) is $||\mathbf{D}(\mathbf{u})||_{L2}$= 3.34 $10^{-3}$ and the CPU time (2.5MHz-4Gb) for the simulation is 247.20 s.

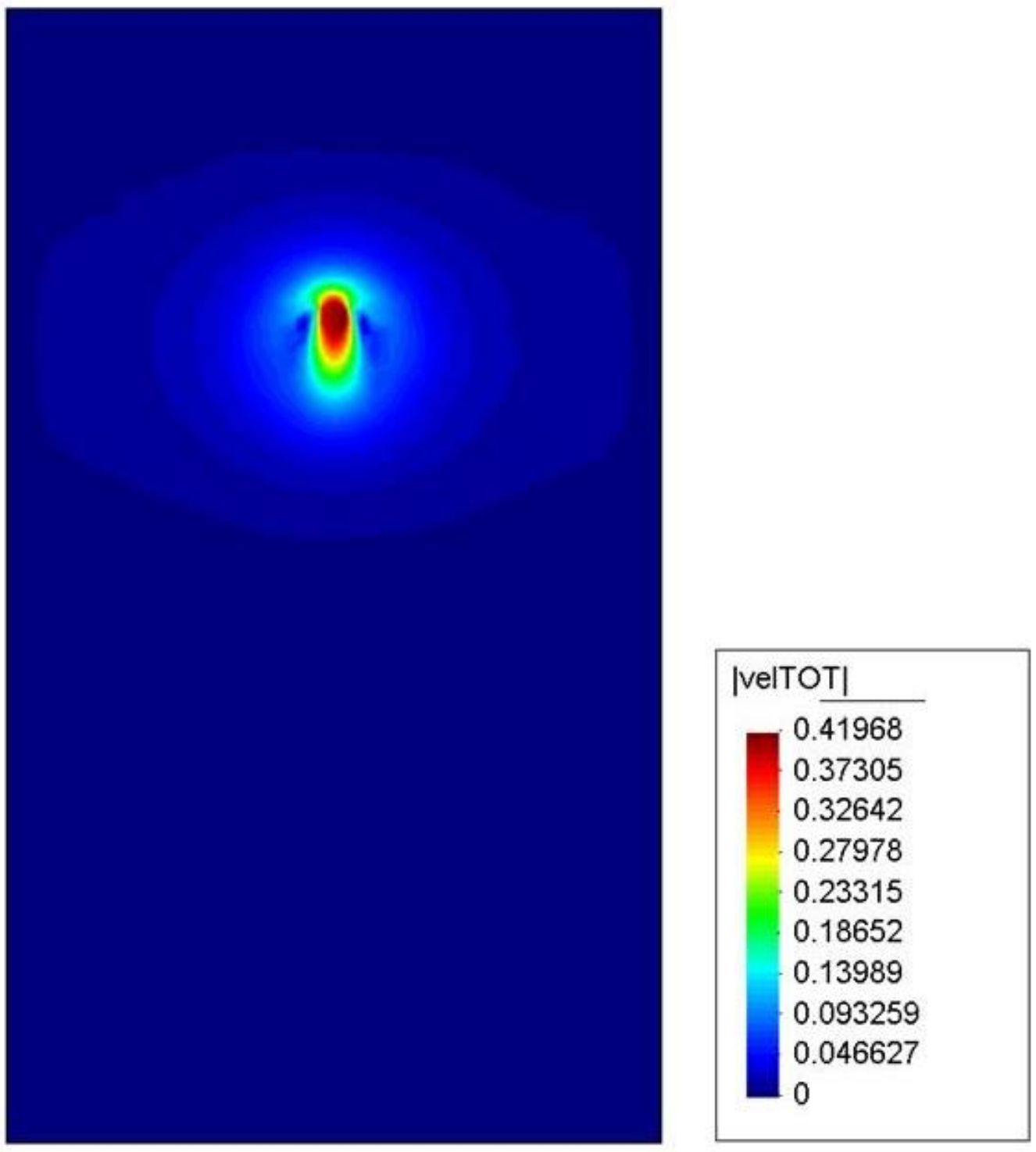

Figure 35: Settling of a circular cylinder with $\rho_s < \rho_f$: velocity field for t = 0.55 s

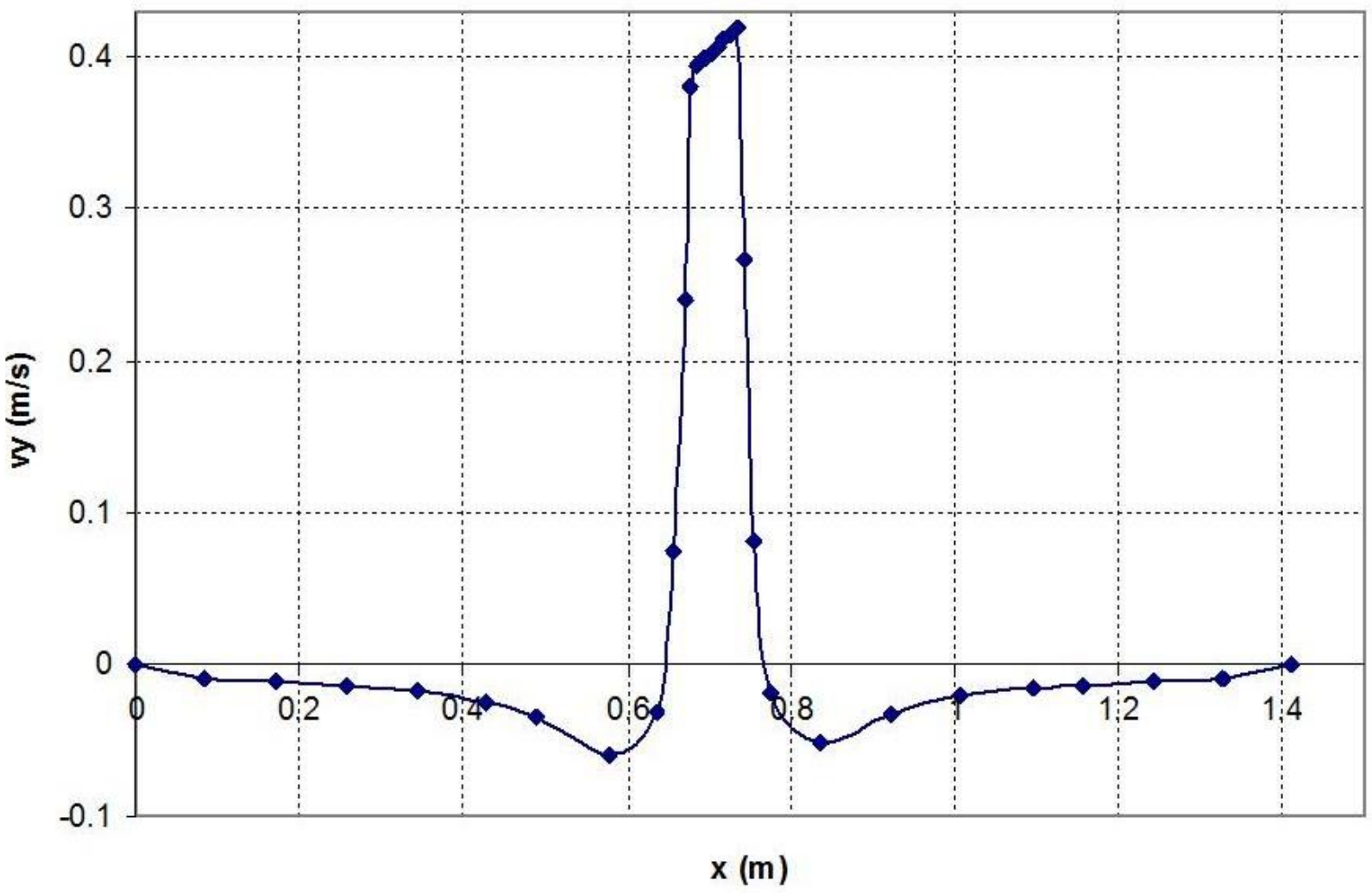

Figure 36: Settling of a circular cylinder with $\rho_s < \rho_f$: vertical velocity distribution along a diametric horizontal line across the domain for t = 0.55 s

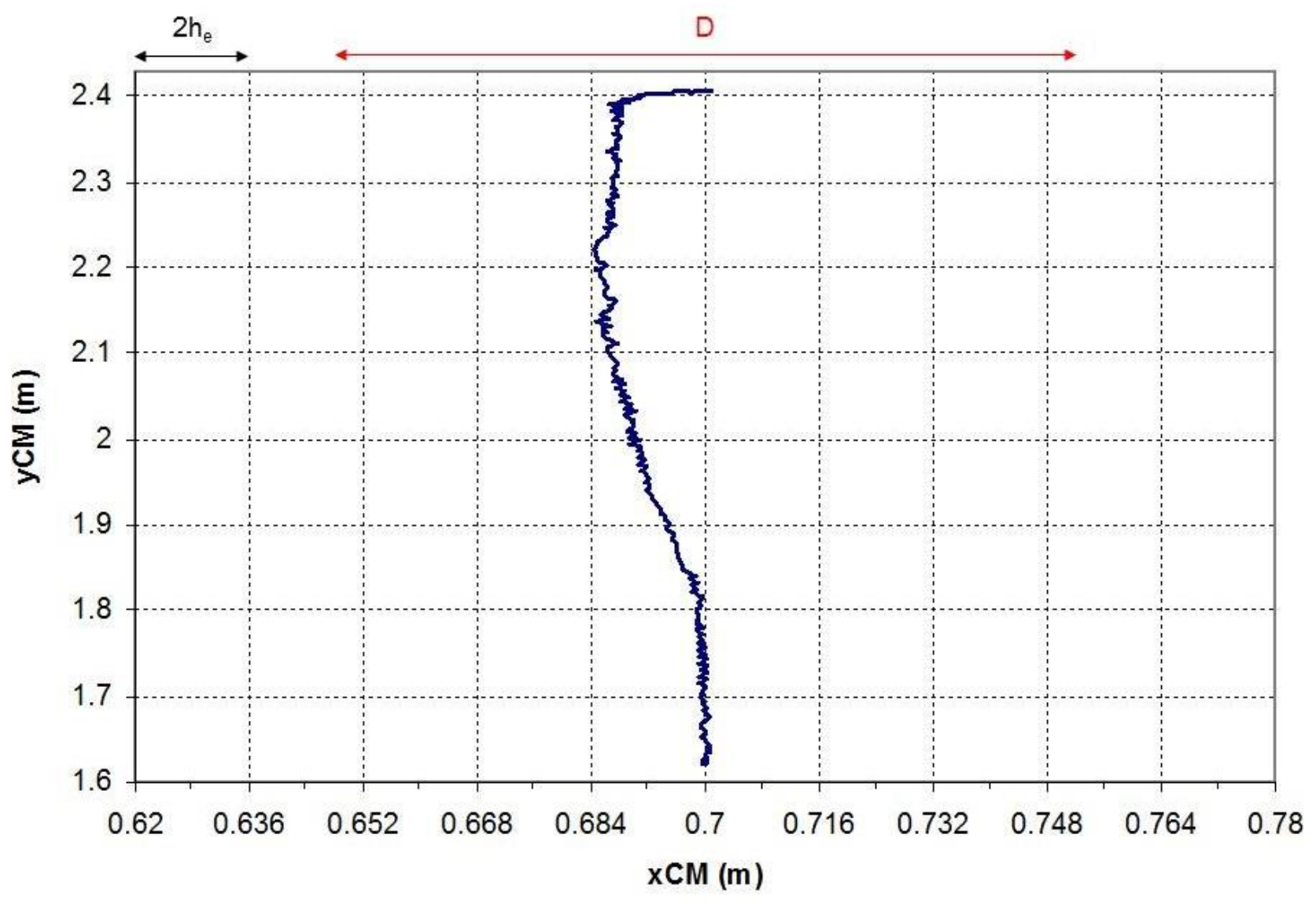

Figure 37: Settling of a circular cylinder with $\rho_s < \rho_f$: path followed by the solid center of mass

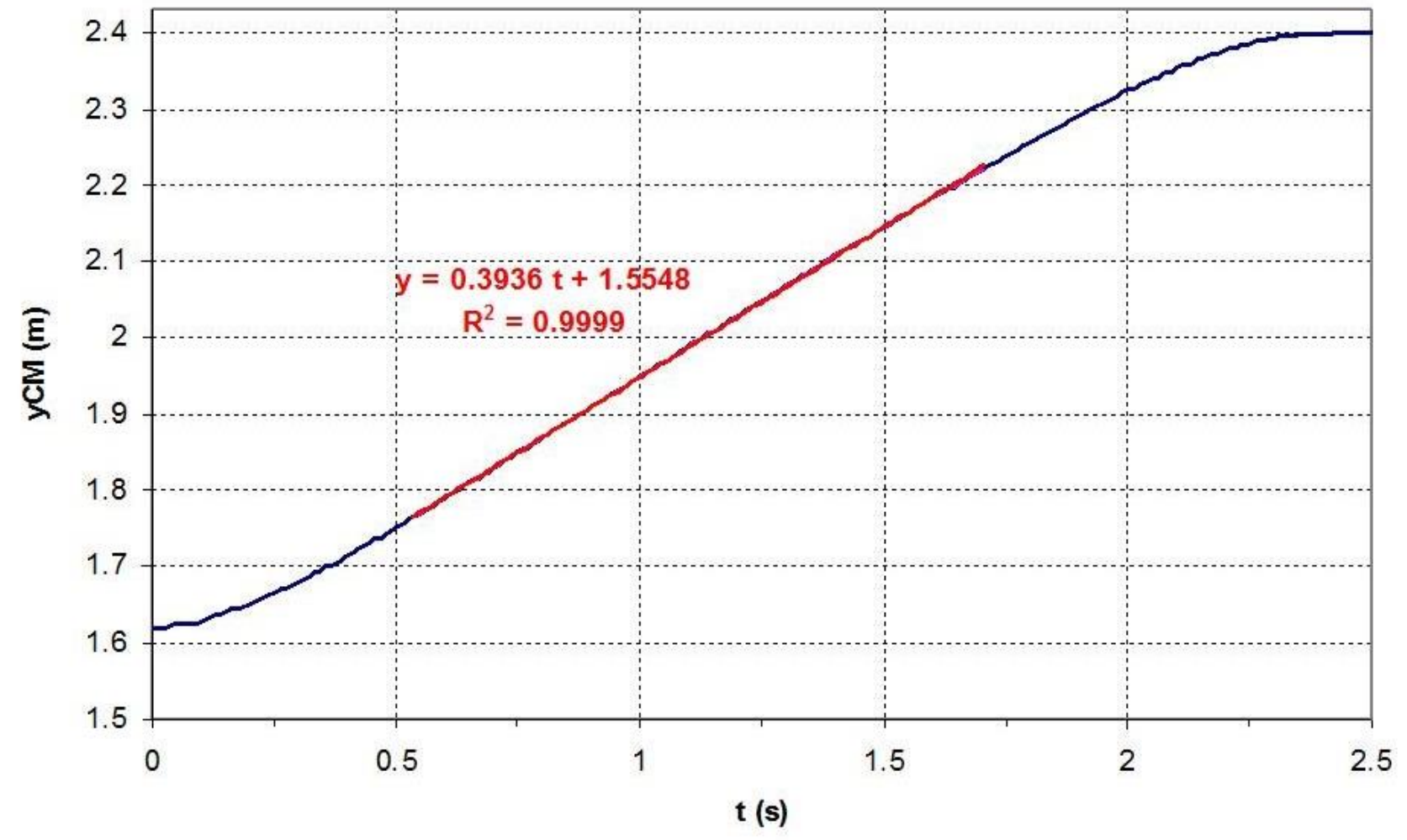

Figure 38: Settling of a circular cylinder with $\rho_s < \rho_f$: evolution in time of the y-coordinate of the solid center of mass and linear regression.

*4.3.3. Settling of a circular cylinder with $\rho_s = \rho_f$*

Finally, in order to assess the stability of the model, the same problem is solved for a solid cylinder whose density is $\rho_s = \rho_f = 1200$ kg/m$^3$. The problem geometry and initial and boundary conditions are the same as in the previous cases (Figures 29-30). The fluid viscosity $\mu = 4$ Pa s and gravity g = 9.8 m/s$^2$. Same non-structured mesh of Figure 30 is used for the computation. The time-step used for the calculation is $\Delta t = 10^{-3}$ s.

No movement restrictions are imposed to the rigid body, and thus it is free to move in all directions. Figure 39 depicts the path followed by the solid center of mass during 3 s (3000 iterations). Figure 40 shows the evolution of its y-coordinate in time. The solid does not either move or distort during the computation, which is a consequence of the absence of numerical instabilities in both velocity and pressure.

After 3 s of simulation and 3 $10^3$ iterations, the calculated error given by (49) is $||\mathbf{D}(\mathbf{u})||_{L2}$= 3.19 $10^{-12}$ and the CPU time (2.5GHz-4Gb) for the simulation is 45.75 s.

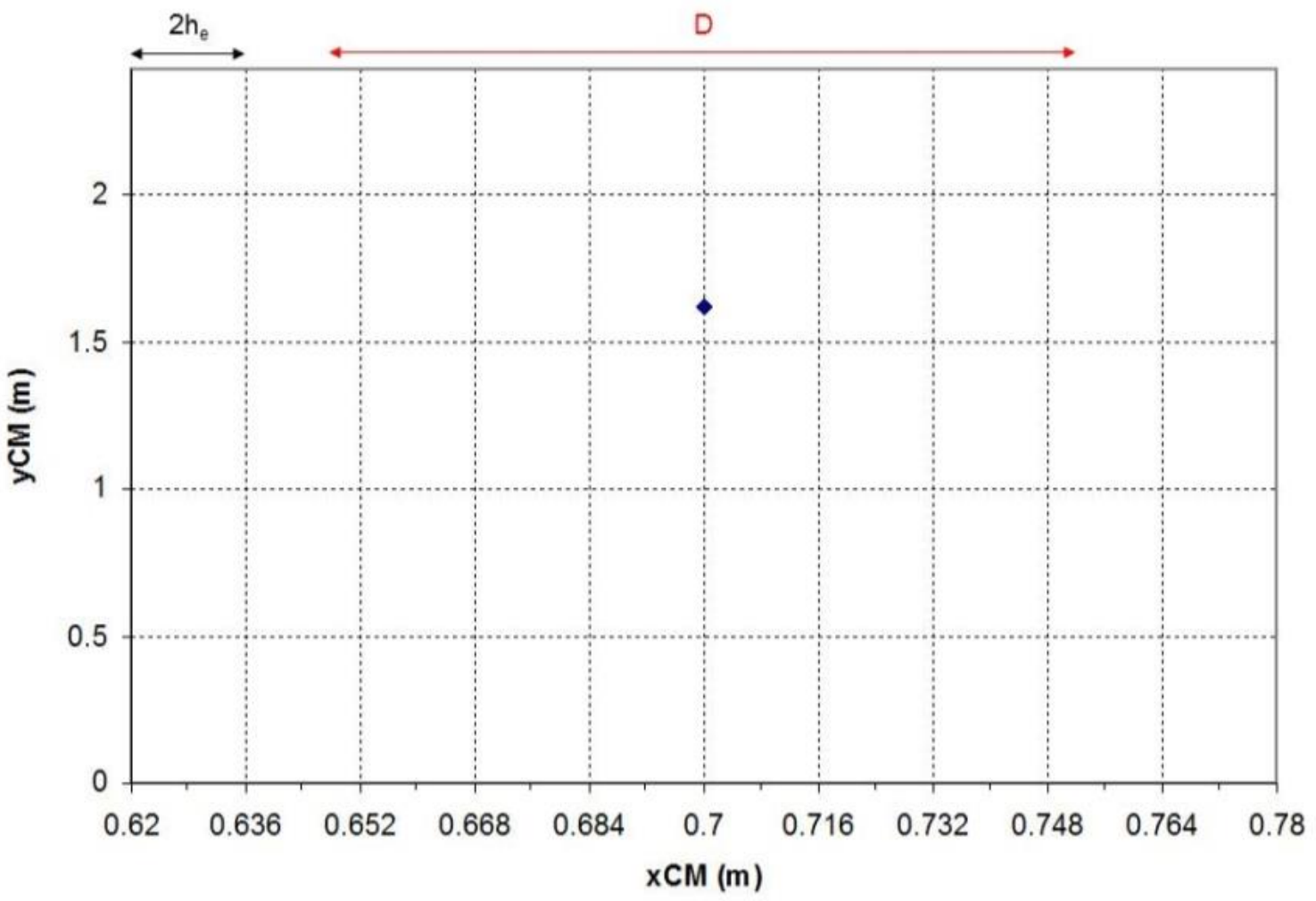

Figure 39: Settling of a circular cylinder with $\rho_s = \rho_f$: path followed by the solid center of mass

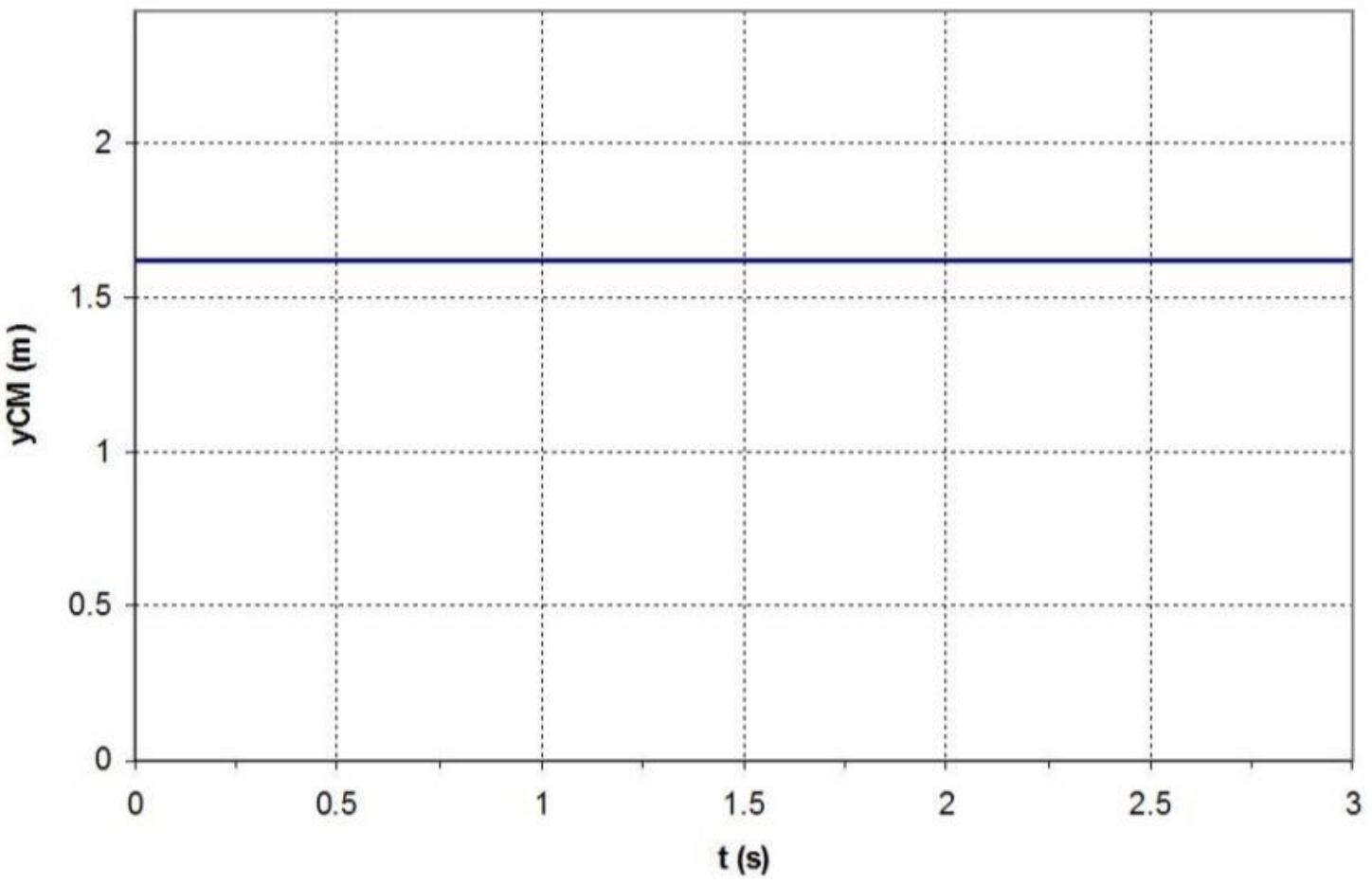

Figure 40: Settling of a circular cylinder with $\rho_s = \rho_f$: evolution in time of the y-coordinate of the solid center of mass.

## 5. Conclusions

A new model for the numerical simulation of a rigid body moving in a viscous fluid flow using FEM has been presented. The model is based on Patankar's projection method [15] combined with the level set algorithm for the tracking of the fluid-solid interface [16,57-61]. The Navier-Stokes equations for a newtonian incompressible fluid are solved using a fractional-step procedure [44,45]. To avoid distortion of the fluid-solid interface the two-step Taylor-Galerkin algorithm is proposed for the solution of the level set advection and correction equations [45-56].

A sensitivity analysis of the method, with respect to mesh size, time step and penalty parameter, has been accomplished, showing the accuracy and efficiency of the method. The distortion of the solid boundary has been quantified, proving that the proposed model preserves the volume and shape of the solid, avoiding numerical deformation of the rigid body. Computation times have been analysed, showing the low computational cost required by the method.

With the validation examples presented in this paper it has been proved that with a reduced number of nodes and at a very low computational cost, the model is able to capture, with a reasonable accuracy, the main features of the flow (i.e. vortex shedding street formation, rear wake detachment angle, dependency of the Strouhal number on the Reynolds number…). The model allows proper calculation of the drag coefficient and it accurately reproduces the dynamics of a solid moving inside a viscous fluid, conserving its volume and shape.

In all tested cases, the numerical results have shown to be in good agreement with other empirical solutions, experimental data and numerical simulations found in the literature [73-84,89], showing the potential of the proposed model as a valuable tool for the numerical analysis of the fluid-solid interaction.

## Acknowledgements

The first author would like to gratefully acknowledge CSIC's financial support for i-LINK project LINKA20203 and the Spanish State Research Agency's (AEI) for project MDM-2017-0737.

## Data availability

The data that support the findings of this study are available from the corresponding author upon reasonable request.